%% file: analytics_cyclotron_s1.0.tex
\newcommand{\printstyle}{reprint}
\newcommand{\halfwidth}{\columnwidth} 

\documentclass[aip,pop,floatfix,\printstyle ,nofootinbib]{revtex4-1}

\input{header.tex}

\begin{document}

\title{Analytic stability boundaries for compressional and global \Alfven eigenmodes driven by fast ions. I. Interaction via ordinary and anomalous cyclotron resonances.}
\author{\myname}
\email{jlestz@pppl.gov}
\affiliation{\Princeton}
\affiliation{\PPPL}
\author{\Nikolai}
\affiliation{\PPPL}
\author{\Elena} 
\affiliation{\PPPL}
\author{\Shawn}
\affiliation{\UCLA}
\author{\Neal}
\affiliation{\UCLA}
\date{\today}
\begin{abstract}
Conditions for net fast ion drive are derived for beam-driven, sub-cyclotron compressional (CAE) and global (GAE) \Alfven eigenmodes, such as those routinely observed in spherical tokamaks such as NSTX(-U) and MAST. Both co- and counter-propagating CAEs and GAEs are investigated, driven by the ordinary and anomalous Doppler-shifted cyclotron resonance with fast ions. Whereas prior results were restricted to vanishingly narrow distributions in velocity space, broad parameter regimes are identified in this work which enable an analytic treatment for realistic fast ion distributions generated by neutral beam injection. The simple, approximate conditions derived in these regimes for beam distributions of realistic width compare well to the numerical evaluation of the full analytic expressions for fast ion drive. Moreover, previous results in the very narrow beam case are corrected and generalized to retain all terms in $\omeganorm$ and $\krat$, which are often assumed to be small parameters but can significantly modify the conditions of drive and damping when they are non-negligible. Favorable agreement is demonstrated between the approximate stability criterion, simulation results, and a large database of NSTX observations of cntr-GAEs. 
\end{abstract}
\maketitle
\section{Introduction}
\label{sec:introduction}

A mixture of high frequency compressional (CAE) and global (GAE) \Alfven eigenmodes were commonly observed on the spherical tokamaks NSTX(-U)\cite{Fredrickson2001PRL,Fredrickson2002POP,Fredrickson2004POP,Fredrickson2006POP,
Crocker2011PPCF,Fredrickson2012NF,Fredrickson2013POP,Crocker2013NF,Crocker2017NF,Fredrickson2018NF,Fredrickson2019POP} and MAST.\cite{Appel2008PPCF,Sharapov2014PP,McClements2017PPCF} These modes may propagate either in the same direction as the plasma current and neutral beam injection (co-propagating) or opposing it (cntr-propagating). They generally have frequencies in the range $\omeganorm = 0.3$ to above the ion cyclotron frequency for CAEs and $\omeganorm = 0.1 - 0.5$ for GAEs with toroidal mode numbers $\abs{n} = 3 - 12$. Dedicated experiments on the large aspect ratio tokamak DIII-D have also observed AE activity in this frequency range,\cite{Heidbrink2006NF,Tang2018APS,Crocker2018APS} allowing comparison between their excitation properties across these different configurations.  

The CAE and GAE, respectively, correspond to compressional (fast magnetosonic) and shear branches of the MHD waves. In a cold, uniform plasma, they have dispersion $\omega = k\va$ and $\omega = \abs{\kpar}\va$, where $\va = B/\sqrt{\mu_0 n_i  m_i}$ is the \Alfven speed. In realistic toroidal geometries with spatial inhomogeneities, the CAE will become localized in the magnetic well in a standing wave configuration\cite{Gorelenkova1998POP,Kolesnichenko1998NF,Smith2003POP,Smith2009PPCF,Smith2017PPCF} with the spectrum of eigenmodes depending on the details of the magnetic geometry.\cite{Mahajan1983bPF,Gorelenkov2002POP,Gorelenkov2002NF,Gorelenkov2006NF} Likewise, the shear \Alfven dispersion becomes spatially dependent in a non-uniform plasma, and modes within this continuum of solutions become strongly damped due to phase mixing.\cite{Heidbrink2008POP,Gorelenkov2014NF} The global \Alfven eigenmode exists below a minimum in the \Alfven continuum (or also possibly above a maximum in the case of nonconventional GAEs\cite{Kolesnichenko2007POP}) where it can avoid the strong continuum damping that would render its excitation more difficult. The discrete spectrum of GAEs exists due to coupling to the CAE, an equilibrium current, current density gradient, and finite $\omeganorm$ effects.\cite{Appert1982PP,Mahajan1983PF,Mahajan1984PF,Li1987PF,Fu1989PF,VanDam1990FT} Excitation of CAEs/GAEs requires a resonant population of energetic particles with sufficient velocity space gradients to overcome damping on the background plasma. The analysis of this paper focuses on fast ions interacting with CAEs/GAEs through the ordinary or anomalous cyclotron resonances. Drive/damping due to the Landau resonance is treated in the second part\cite{Lestz2019p2} of this series. 

The analytic study of the conditions for excitation of CAEs and GAEs is motivated by their experimental observations across many devices. The presence of these modes has been linked to anomalous electron energy transport in NSTX,\cite{Stutman2009PRL,Ren2017NF} which may be explained by orbit stochastization\cite{Gorelenkov2010NF} and energy channeling at the \Alfven resonance location.\cite{Kolesnichenko2010PRL,Kolesnichenko2010NF,Belova2015PRL,Belova2017POP,Kolesnichenko2018POP,Kolesnichenko2018bPOP} During early operations of NSTX-U, robust stabilization of GAEs by the addition of a small amount of power in the new off-axis neutral beam source was discovered and subsequently reproduced with numerical modeling and analytic theory.\cite{Fredrickson2017PRL,Fredrickson2018NF,Kaye2019NF,Belova2019POP} Further understanding of these processes will be aided by the new stability conditions derived here. 

General expressions for the growth rate of these instabilities were originally derived for mono-energetic beam\cite{Belikov1968ZHETF,*Belikov1969JETP} and bi-Maxwellian\cite{Timofeevv5} distributions, as well as for an arbitrary distribution\cite{Mikhailovskiiv6} in a uniform plasma. These derivations were later extended and applied to NBI-driven CAEs/GAEs in various experimental conditions dating back to the TFTR era\cite{Gorelenkov1995POP,Gorelenkov1995NF} and continuing in more recent years with applications to JET\cite{Gorelenkov2002bNF} and NSTX.\cite{Gorelenkov2003NF,Kolesnichenko2006POP} The recent studies on NBI-driven modes had two key limitations. First, they did not correctly treat the cutoff at the injection energy, an approach suitable for shifted Maxwellians generated by ICRF, but not for slowing down distributions from NBI. Second, they assumed a delta function in pitch for tractability, which is unrealistic considering the more broad distributions present in experiments, as inferred from Monte Carlo codes such as the \NUBEAM\cite{Pankin2004CPC} module in \TRANSP.\cite{Goldston1982JCP} Prior studies also assume $\kpar \ll \kperp$ and $\omega \ll \omegaci$ as simplifying approximations, whereas the modes excited in spherical tokamaks such as NSTX may have frequencies approaching $\omega \lesssim \omegaci$ and $\kpar \like \kperp$.

The derivation presented in this paper corrects and builds on prior work by providing a local expression for the fast ion drive due to a general beam-like distribution interacting via the ordinary and anomalous cyclotron resonances. The effect of finite injection energy of NBI distributions is included consistently, yielding a previously overlooked instability regime. Terms to all order in $\omeganorm$ and $\krat$ are kept for applicability to the entire possible spectrum of modes. As in previous works, full finite Larmor radius (FLR) terms are also retained. The analytic expression can be integrated numerically for any chosen parameters in order to determine if the full fast ion distribution is net driving or damping. More interestingly, it is found that when the beam is sufficiently wide in velocity space, such as realistic distributions resulting from NBI, the integral can be evaluated approximately in terms of elementary functions, yielding compact conditions for net fast ion drive/damping that depend only on a small set of parameters describing the fast ion and mode parameters. Such expressions grant new insights into the spectrum of CAEs and GAEs that may be excited by a given fast ion distribution, as well as providing intuition for interpreting experimental results. Since damping sources such as electron Landau and continuum damping are not addressed in this work, the net fast ion drive conditions derived here should be considered as necessary but not sufficient conditions for instability. 

The paper is structured as follows. The dispersion relations, resonance condition, and model fast ion distribution function used in this paper are described in \secref{sec:dispres}. In \secref{sec:derivation}, the local analytic expression for the CAE and GAE growth rates is adapted from \citeref{Mikhailovskiiv6} and applied to the fast ion distribution of interest. Approximations are applied to this expression in \secref{sec:approxstab} in order to derive useful instability criteria for the cases of a very narrow beam width in velocity space (\secref{sec:narrow}) and a beam with realistic width (\secref{sec:wide}) when FLR effects are small (\secref{sec:slow}) and large (\secref{sec:fast}). The derived conditions are also compared against the numerically calculated growth rates for realistic parameter values in \secref{sec:approxstab}. In \secref{sec:stabao}, the dependence of the fast ion drive/damping on the mode properties ($\omeganorm$ and $\krat$) is presented and compared against conclusions drawn from the approximate stability boundaries. A comparison of the approximate stability conditions against a database of cntr-GAE activity in NSTX are shown in \secref{sec:expcomp}. Lastly, a summary of the main results and discussion of their significance is given in \secref{sec:summary}.

\section{Dispersion, Resonance Condition, and Fast Ion Distribution} 
\label{sec:dispres}

One goal of this paper is to extend previous derivations to include finite $\omeganorm$ and $\krat$ effects in the stability calculation, since experimental observations and modeling of NSTX suggests that these quantities may not always be small. Experimental observations often show CAEs with frequencies from $\omeganorm = 0.3$ to exceeding the cyclotron frequency. GAEs are observed with somewhat lower frequencies of $\omeganorm \approx 0.1 - 0.5$. While $\kperp$ can not be measured accurately on NSTX due to limited poloidal coil resolution, it can be calculated for the most unstable modes excited in simulations,\cite{Belova2017POP} which show that $\krat \approx 1$ is not uncommon, and can even reach $\krat > 3$ in some cases. This motivates using the full, unsimplified dispersion relations in uniform geometry when numerically calculating the growth rate, instead of using the common $\omeganorm\ll 1$ and $\krat \ll 1$ assumptions found in previous works. The more complicated eigenmode equations in nonuniform toroidal systems\cite{Appert1982PP,Mahajan1983PF,Mahajan1983bPF,Gorelenkov1995NF,Gorelenkov1995POP} have been derived in the past but are too complicated for our purposes. 

Define $\omegabar = \omega/\omegacio$, $N = k\va/\omega$, $A = (1 - \omegabar^2)^{-1}$, and also $F^2 = \kpar^2/k^2$, $G = 1 + F^2$. Here, $\omegacio$ is the on-axis ion cyclotron frequency. Then in uniform geometry, the local dispersion in the MHD limits of $E_\parallel \ll E_\perp$ and $\omega \ll \abs{\omegace},\omegape$ is readily given by\cite{Stix1975NF} 

\begin{equation}
N^2 = \frac{AG}{2F^2}\left[1 \pm \sqrt{1 - \frac{4F^2}{AG^2}}\right]
\label{eq:stixdisp}
\end{equation}

The ``$-$" solution corresponds to the compressional \Alfven wave (CAW), while the ``$+$" solution corresponds to the shear \Alfven wave (SAW). The coupled dispersion in \eqref{eq:stixdisp} will be used in the full analytic expression for fast ion drive. Notably, it can modify the polarization of the two modes, which in turn changes how the finite Larmor radius (FLR) effects from the fast ions contribute to the growth rate (see \eqref{eq:Jlm}). Its low frequency approximations are $\omega \approx k\va$ for CAWs and $\omega \approx \abs{\kpar}\va$ for SAWs. Throughout the paper, CAW/CAE and SAW/GAE will be used interchangeably, where CAW and SAW formally refer to the solutions in a uniform slab, while CAE and GAE refer to their analogues in nonuniform and bounded geometries. Net energy transfer between a mode and the fast ions requires a sub-population of particles obeying the Doppler-shifted cyclotron resonance. 

\begin{align}
\omega - \avg{\kpar\vpar} - \avg{\kperp\vdrift} = \lres\avg{\omegaci}
\label{eq:rescon}
\end{align}

Here, $\avg{\dots}$ denotes poloidal orbit averaging and $\lres$ is an integer cyclotron resonance coefficient. Two resonances are studied in detail in this work for the sub-cyclotron modes: the $\lres = 1$ ordinary cyclotron resonance and $\lres = -1$ anomalous cyclotron resonance. Orbit averaging in \eqref{eq:rescon} is required to satisfy the global resonance condition, as opposed to the local resonance, which describes a net synchronization condition between the wave and particle on average over its orbit, even while not being in constant resonance at all points in time. This resonance condition is applicable so long as the growth rate of the mode is sufficiently smaller than the inverse particle transit time, which is satisfied by these modes according to \HYM simulations.\cite{Lestz2019sim} 
 
In this paper, we will make the approximation of $\abs{\kperp\vdrift} \ll \abs{\kpar\vpar}$. Consequently, when $\omega < \omegaci$ and $\avg{\vpar} > 0$ (co-injection), \eqref{eq:rescon} can only be satisfied for $\lres = 1$ if $\kpar < 0$ (mode propagates counter to the fast ions). Likewise, $\lres = -1$ requires $\kpar > 0$, corresponding to co-propagation. Due to periodicity, the drift term can be approximated for passing particles\cite{Belikov2003POP} as $\avg{\kperp\vdrift} \approx s\avg{\vpar}/qR$ for integer $s$, though this term yields relatively small corrections due to the large values of $\abs{\kpar}$ relevant to these modes. In this approximation, the resonance condition can be rewritten as $\omega - \kpars\vpres = \lres\avg{\omegaci}$ with $\kpars = \kpar + s/q R$. Conversely, for trapped particles the drift term can be approximated as\cite{Belikov2004POP} $\avg{\kperp\vdrift} \approx s\omega_b$. Previous \HYM simulations indicate that the $s = \pm 1$ sidebands are usually more relevant than larger $\abs{s}$.\cite{Belova2017POP} For quantitatively accurate growth rates, all sidebands should be summed over, as done in \citeref{Gorelenkov1995POP} in the limit of $\omega\gtrsim\omegaci\gg\omega_b$, and also in \citeref{Kolesnichenko2006POP}. Practically, these procedures require complicated non-local calculations which would preclude analytic progress except in extraordinarily special cases, contrary to the purpose of this work, which is to derive broadly applicable instability conditions. To this end, only the primary resonance $(s = 0)$ will be kept when deriving approximate stability boundaries in \secref{sec:approxstab}.  

Combination of the resonance condition with approximate dispersion relations can yield relations that will be useful later on. Introduce $\omegacires \defined \avg{\omegaci}/\omegacio$ as the average cyclotron frequency of the resonant particles, normalized to the on-axis cyclotron frequency $\omegacio$. This value is approximately 0.9, as inferred from inspection of the resonant particles in relevant \HYM simulations. Then defining $\vpres \defined \avg{\vpar} > 0$ (treating co-injected particles only) and rearranging \eqref{eq:rescon} gives

\begin{align}
\frac{\vpres}{\va} &= \abs{\frac{\omega}{\kpar\va}}\abs{1 - \frac{\lres\omegacires}{\omegabar}} \\ 
\label{eq:vpres}
&\approx \left\{\begin{array}{ll}
\abs{1 - \frac{\lres\omegacires}{\omegabar}} & \GAElab \\ 
\sqrt{1 + \frac{\kperp^2}{\kpar^2}}\abs{1 - \frac{\lres\omegacires}{\omegabar}} & \CAElab
\end{array}\right.
\end{align}

The stability calculation will be applied to a slowing down, beam-like background distribution of fast ions, motivated by theory and \NUBEAM modeling of NSTX discharges.\cite{Belova2017POP} In order to satisfy the steady state Vlasov equation, the distribution is written as a function of constants of motion $v = \sqrt{2\W/m_i}$ and $\lambda = \mu B_0 / \W$ in separable form: $\fb(v,\lambda)= C_f n_b f_1(v)f_2(\lambda)$, defined below

\begin{subequations}
\begin{align}
\label{eq:F1}
f_1(v) &= \frac{\ftail(v;v_0)}{v^3 + v_c^3} \\ 
\label{eq:F2}
f_2(\lambda) &= \exp\left(-\left(\lambda - \lambda_0\right)^2 / \Delta\lambda^2\right)
\end{align}
\label{eq:Fdistr}
\end{subequations}

The constant $C_f$ is for normalization.  
The first component $f_1(v)$ is a slowing down function in energy with a cutoff at the injection energy $v_0$ and a critical velocity $v_c$. The cutoff at $v = v_0$ is contained within $\ftail(v;v_0)$, which is in general a function which rapidly goes to zero for $v>v_0$. For ease of calculation, this is assumed to be a step function. The second component $f_2(\lambda)$ is a Gaussian distribution centered on some central value $\linj$ with width $\dl$. The variable $\lambda$ is a trapping parameter. To lowest order in $\mu \approx \mu_0$, it can be re-written as $\lambda = (\vperp^2/v^2)(\omegacio/\omegaci)$. Then, assuming a tokamak-like field $B \approx B_0/(1 + \epsilon\cos\theta)$ for $\epsilon = r/R$, passing particles will have $0 < \lambda < 1 - \epsilon$ and trapped particles will have $1 - \epsilon < \lambda < 1 + \epsilon$. Loosely, smaller $\lambda$ means the particle's velocity is more field aligned, such that $\lambda$ is a complementary variable to a particle's pitch $\vpar/v$. For analytic tractability, $\linj$ and $\dl$ are treated as constants in this model, ignoring any velocity dependence of these parameters which may be present, especially broadening in $\lambda$ at lower energies due to pitch angle scattering. The dependence on $\pphi$, is neglected in this study for simplicity, as it is expected to be less relevant for the high frequencies of interest for these modes.   

The NSTX operating space spanned a range of normalized injection velocity $\vinj = 2 - 6$, depending on the beam voltage (typically $60 - 90$ keV at $2 - 6$ MW) and field strength ($0.25 - 0.50$ T) for each discharge. The central trapping parameter $\linj$ and beam width $\dl$ are mostly determined by the neutral beam's geometry and collimation, yielding typical $\linj = 0.5 - 0.7$ and $\dl = 0.3$. For this study, $v_c = v_0/2$ is used as a characteristic value. The new beam line on NSTX-U has much more tangential injection, with $\linj \approx 0$, and also lower $\vinj = 1 - 3$ due to higher nominal field strength. A comparison between the model fast ion distribution and a \NUBEAM calculation for the well-studied H-mode discharge $\# 141398$, can be found in Fig. 5 of \citeref{Belova2003POP}. 

\section{Fast Ion Drive for General Beam Distribution in the Local Approximation} 
\label{sec:derivation}

In this section, the fast ion drive/damping is derived perturbatively in the local approximation for a two component plasma comprised of a cold bulk plasma and a minority hot ion kinetic population, and applied to the general beam distribution of interest. The formula presented here extends the results obtained in \citeref{Gorelenkov2003NF,Kolesnichenko2006POP}, which focused on $\omega \ll \omegaci$, $\kpar \ll \kperp$, and also did not study high frequency co-propagating  modes ($\lres = -1$ cyclotron resonance coefficient). In contrast, the following derivation is appropriate for all values of $\omeganorm$ and $\krat$, which is important since mode frequencies can be on the order $\omeganorm \like 0.5$ or larger, and in contrast to the common large tokamak assumption, $\krat$ can be of order unity, as inferred from simulations.\cite{Lestz2018POP,Lestz2019sim} 

\subsection{Derivation}
\label{sec:subderiv}

The general dispersion is given by 

\begin{equation}
\abs{\epsilon_{ij} - n^2\left(\delta_{ij} - \frac{k_i k_j}{k^2}\right)} = 0 
\end{equation}

Here, $n = k c / \omega$ is the index of refraction, $\epsilon_{ij} = \delta_{ij} + \sum_s \epsilon_{ij}^s$ is the dielectric tensor. Without loss of generality, assume $\vB_0 = B_0\hat{z}$ and $\vk = \kpar\hat{z} + \kperp\hat{x}$. Then the dispersion is determined by 

\begin{equation}
\left(\begin{array}{cc} 
\epsilon_{11} - n_\parallel^2 & \epsilon_{12} \\ 
\epsilon_{21} & \epsilon_{22} - n^2
\end{array}\right)
\left(\begin{array}{cc}
E_x \\ E_y 
\end{array}\right) = 0
\end{equation}

The rest of the components are irrelevant in the MHD regime where $E_z \ll E_x,E_y$. For the cold bulk components,

\begin{equation}
\delta_{ij} + \epsilon_{ij}^{th,e} + \epsilon_{ij}^{th,i} = 
\left(\begin{array}{cc}
S & -iD \\ 
iD & S
\end{array}\right) 
\end{equation}

Above, $S = 1 - \sum_s \omegaps^2/(\omega^2 - \omegacs^2)$ and $D = \sum_s \omegacs\omegaps^2/(\omega(\omega^2 - \omegacs^2))$, where $\omegaps = \sqrt{n_s q_s^2/(m_s \epsilon_0)}$ and $\omegacs = q_s B_0 / m_s$ are the plasma frequency and signed cyclotron frequency for each species $s$. When $\omega\ll \omegape,\abs{\omegace}$, we can approximate $S \approx A c^2/\va^2$ and $D \approx -\omegabar A c^2/\va^2$, where as earlier $A = 1/(1 - \omegabar^2)$ and $\omegabar = \omega/\omegacio$. Setting $\Kij = \va^2 \epsilon_{ij}^b/c^2$ and also defining $y = \omega^2/(k^2\va^2) = N^{-2}$, the full dispersion is given by 

\begin{multline}
\left(y - F^2 \bvar - y \bvar \Kxx\right)\left(y - \bvar - y \bvar \Kyy\right) \\ 
- y^2\left(\omegabar + \bvar\Kxy\right)^2 = 0 
\end{multline}

Neglecting the fast ion component (setting $\Kij = 0$) recovers the MHD dispersion in \eqref{eq:stixdisp}. Letting $\omega = \omega_0 + \omega_1$ with $\omega_1 \ll \omega_0$ and solving perturbatively to first order in $\Kij \like n_b/n_e \ll 1$ yields the growth rate as

\begin{align}
\frac{\omega_1}{\omega_0} = \frac{y_0\left[\Kxx(y_0 - \bvar_0) -2\omegabar_0 y_0\abs{\Kxy} + (y_0 - F^2 \bvar_0)\Kyy\right]}{2\left(y_0^2 - F^2\right)}
\label{eq:omegapert}
\end{align}

As defined in \secref{sec:dispres}, $F^2 = \kpar^2/k^2$. All quantities with subscript $0$ are understood to be evaluated using $\omega = \omega_0$, \ie the unperturbed frequency given by \eqref{eq:stixdisp}. The tensor elements $\Kij$ can be calculated from Eq. A24 in \citeref{Mikhailovskiiv6}:

\begin{align}
\label{eq:ktens}
\Kij &= \frac{n_b}{n_e}\frac{\omegaci^2}{\omega}\int \vperp d\vperp d\vpar \sum_{\lres = -\infty}^\infty \frac{\vperp^2 g_{ij}^\lres(\flr)}{\omega - \kpar\vpar - \lres\omegaci}\hat{\pi}\fb \\ 
\label{eq:piparperp}
\text{where } \hat{\pi} &= \frac{1}{\vperp}\pderiv{}{\vperp} + \frac{\kpar}{\omega}\left(\pderiv{}{\vpar} - \frac{\vpar}{\vperp}\pderiv{}{\vperp}\right) \\ 
g_{ij}^\lres(\flr) &= \left(\begin{array}{cc}
\lres^2 \Jl^2/\flr^2 & i\lres \Jl^\prime \Jl/\flr \\ 
-i\lres \Jl^\prime \Jl/\flr & (\Jl^\prime)^2
\end{array}\right),\, \flr = \kperp\rhob
\end{align}

Above, $\rhob = \vperp/\omegaci$ is the Larmor radius of the fast ions, and the distribution is normalized such that $\int \vperp \fb d\vperp d\vpar = 1$. The finite Larmor radius (FLR) effects from the fast ions are contained in $g_{ij}^\lres(\flr)$, with $\Jl(\flr)$ denoting the $\lres^{th}$ order Bessel function of the first kind. In order to keep only the resonant contribution to the growth rate, we make the formal transformation $(\omega - \kpar\vpar - \lres\omegaci)^{-1} \rightarrow -i\pi\delta(\vpar - \vpreslres)/\abs{\kpar}$ with $\vpreslres = (\omega - \lres\omegaci)/\kpar$ the parallel velocity of the resonant fast ions. Then substituting \eqref{eq:ktens} into \eqref{eq:omegapert} and identifying the growth rate $\gamma = \text{Im}(\omega_1)$,  

\begin{align*}
\label{eq:gammageneral}
&\frac{\gamma}{\omegaci} = \frac{\pi}{2}\frac{n_b}{n_e}\sum_\lres \abs{\frac{\vpreslres}{\omegabar - \lres}} \\ 
&\quad\quad\times \int d\vperp d\vpar \vperp^3 \delta(\vpar - \vpreslres)\hat{\pi}_\lres\fb \Jlm(\flr) \numberthis \\ 
&\text{where } \hat{\pi}_\lres = \frac{1}{\W}\left[\left(\frac{\lres}{\omegabar} - x\right)\pderiv{}{x} + \frac{v}{2}\pderiv{}{v}\right] \numberthis
\end{align*} 

The variable $x = \vperp^2/v^2 = \lambda\omegacires$ was introduced so that the gradients $\hat{\pi}\fb$ can be re-written in the natural coordinates of the distribution. Note that $\Jlm(\flr)$ is the ``FLR function" for cyclotron resonance $\lres$ and mode $m$ (= `$C$' for CAE and `$G$' for GAE), defined as 
\begin{align}
\Jlm(\flr) &\defined \frac{y_0}{y_0^2 - F^2}\left[\sqrt{y_0 - \bvar_0}\frac{\lres\Jl}{\flr} \mp \sqrt{y_0 - F^2 \bvar_0}\deriv{\Jl}{\flr}\right]^2
\label{eq:Jlm}
\end{align}

Above, the ``$-$" corresponds to CAEs and the ``$+$" for GAEs. Defining $\alpha = \krat$, the FLR parameter $\flr$ may also be re-written in the following form: 

\begin{align}
\label{eq:zsimp}
\flr &= \kperp\rhob \defined \zp \sqrt{\frac{x}{1-x}} \\ 
\zp &= \frac{\kperp\vpres}{\omegaci} = \frac{\abs{\omegabar - \lres\omegacires}}{\alpha}
\label{eq:zp}
\end{align}

The modulation parameter $\zp$ contains information about the mode characteristics and is a measure of how rapidly the integrand in \eqref{eq:gammageneral} is oscillating. The expression in \eqref{eq:zp} follows from the resonance condition in \eqref{eq:vpres}. The complicated form of $\Jlm(\flr)$ is due to coupling between the pure compressional and shear branches of the dispersion resulting from finite $\omeganorm$ and also modified by finite $\krat$, so it is worthwhile to highlight some of its properties. The FLR function $\Jlm(\flr)$ is non-negative for both modes when $\omeganorm < 1$. For CAEs, $y_0 \geq 1 \geq \bvar_0, F, F^2\bvar_0$ according to \eqref{eq:stixdisp}, so the square root arguments and leading factors are all positive. In contrast, for GAEs, $y_0 \leq \bvar_0, F, F^2\bvar_0$, so the arguments of the square roots as well as the leading factors are all negative, with signs canceling out. 

As a useful example, consider the limit of $\omeganorm \ll 1$. In that case, $y_0 = 1 + \omegabar^2\alpha^2 \plusord{\omegabar^4}$ for CAEs and $y_0 = F^2 - \omegabar^2\alpha^2 \plusord{\omegabar^4}$ for GAEs. Then $\Jlm(\flr)$ simplifies substantially to

\begin{subequations}
\label{eq:Jlmsmallappx}
\begin{align}
\label{eq:Jlmsmallappx-cae}
\lim_{\omegabar\rightarrow 0} \Jlc(\flr) &= \left(\Jlprime\right)^2
\CAElab \\ 
\label{eq:Jlmsmallappx-gae}
\lim_{\omegabar\rightarrow 0} \Jlg(\flr) &= \left\{\begin{array}{ll}
\left(\lres\Jl/\flr\right)^2 
& \lres \neq 0 \\ 
\left(\omegabar\alpha^2 J_1\right)^2 & \lres = 0 
\end{array}\right. 
\GAElab
\end{align}
\end{subequations} 

In another limit, where $0 < \omegabar < 1$ and $\alpha \gg 1$, the dispersion from \eqref{eq:stixdisp} reduces to $y_0 = 1 + \omegabar$ for CAEs and $y_0 = 1 - \omegabar$ for GAEs, simplifying the FLR function to 

\begin{subequations}
\label{eq:Jlmbigappx}
\begin{align}
\label{eq:Jlmbigappx-a}
\lim_{\alpha\rightarrow \infty} \Jlm(\flr) &= \frac{\left(1 \pm \omegabar\right)^2}{2\pm \omegabar}\left(\Jlprime \mp \frac{\lres\Jl}{\flr}\right)^2 \\ 
\label{eq:Jlmbigappx-cae}
\lim_{\alpha\rightarrow \infty} \Jlc(\flr) &= \frac{\left(1 + \omegabar\right)^2}{2 + \omegabar} J_{\lres+1}^2 \CAElab \\ 
\label{eq:Jlmbigappx-gae}
\lim_{\alpha\rightarrow \infty} \Jlg(\flr) &= \frac{\left(1 - \omegabar\right)^2}{2 - \omegabar} J_{\lres-1}^2 \GAElab
\end{align}
\end{subequations}

In \eqref{eq:Jlmbigappx-a}, the top signs are for CAEs, and the bottom signs for GAEs. The forms in \eqref{eq:Jlmsmallappx} match those used in \citeref{Gorelenkov2003NF,Kolesnichenko2006POP} in the same limit, and the limit of $\alpha \rightarrow 0$ of \eqref{eq:Jlm} reproduces the FLR function used in \citeref{Belikov2003POP,Belikov2004POP}. Applying \eqref{eq:gammageneral} to the general beam distribution in \eqref{eq:Fdistr} and defining $\bres_\lres = \vpreslres^2/v_0^2$ yields 

\begin{widetext}
\begin{multline}
\frac{\gamma}{\omegaci} = -\frac{n_b}{n_e}\frac{\pi C_f v_0^3 }{2 v_c^3} \sum_\lres \frac{\bres_\lres^{3/2}}{\abs{\omegabar-\lres}}
\left\{ \int_0^{1-\bres_\lres} \frac{x \Jlm(\flr(x,\zp))}{(1-x)^2}\frac{e^{-(x-\xinj)^2/\dx^2}}{1 + \frac{v_0^3}{v_c^3}\left(\frac{\bres_\lres}{1-x}\right)^{3/2}}
\left[\frac{1}{\dx^2}\left(\frac{\lres}{\omegabar} - x\right)(x-\xinj) + \frac{3}{4}\frac{1}{1 + \frac{v_c^3}{v_0^3}\left(\frac{1-x}{\bres_\lres}\right)^{3/2}}\right]dx \right. \\  
\left.\vphantom{\left[\frac{3}{4}\frac{1}{1 + \left(\frac{1-x}{4\bres_\lres}\right)^{3/2}}\right]}
+ \frac{\bres_\lres^{-1}-1}{2\left(1 + \frac{v_0^3}{v_c^3}\right)}e^{-(1 - \bres_\lres-\xinj)^2/\dx^2}\Jlm\left(\zp\sqrt{\bres_\lres^{-1}-1}\right)\right\}
\label{eq:gammabeam}
\end{multline}
\end{widetext}

The upper integration bound is a consequence of the finite injection energy since $\abs{\vpres} = v\sqrt{1 - x} < v_0 \sqrt{1 - x} \rightarrow x < 1 - \vpres^2/v_0^2$. All quantitative calculations in this paper assume $v_c = v_0/2$ and $n_b/n_e = 5.3\%$, based on the conditions in the well-studied NSTX H-mode discharge $\# 141398$. The normalization constant is given by 

\begin{align}
C_f^{-1} &= \frac{1}{3}\ln\left(1 + \frac{v_0^3}{v_c^3}\right)\int_0^1 \frac{e^{-(x-\xinj)^2/\dx^2}}{\sqrt{1 - x}}dx 
\end{align}

This approach required two large assumptions in order to make the problem tractable. First, a local assumption was made in order to eliminate the spatial integrals, which require knowledge or detailed assumptions about the equilibrium profiles and mode structures, whereas we seek a simple criteria depending only on a few parameters ($\vinj, \linj, \omeganorm, \krat, \lres$) for broad comparison with experimental or simulation results. Hence, all equilibrium quantities in \eqref{eq:gammabeam} are understood to be taken at the peak of the mode structure, generally between the magnetic axis and mid-radius on the low-field side, where CAEs are localized due to a magnetic well and GAEs are localized due to a minimum in the \Alfven continuum. As a consequence, the accuracy of the drive/damping magnitude may be limited, however this approximation should not affect the sign of the expression, so it can still be used to distinguish net fast ion drive vs damping, which is the primary goal of this work. Second, the derivative with respect to $\pphi$ has been neglected in this derivation, which would be important for modes at lower frequencies (\eg for TAEs where it is the main source of drive) or fast ion distributions with very sharp spatial gradients, which is atypical for NBI. 

\subsection{Properties of Fast Ion Drive}
\label{sec:properties}

The expression in \eqref{eq:gammabeam} represents the local perturbative growth rate for CAE/GAEs in application to a general beam-like distribution of fast ions, keeping all terms from $\omeganorm$, $\krat$, and $\kperp\rhob$. The derivation presented in this section has some additional consequences worth highlighting. Observe that only the term in square brackets can change sign since the coefficient in front of the integral will always be negative, and the portions of the integrand not enclosed in square brackets are strictly nonnegative. Hence regions of the integrand where the term in brackets is negative are driving, and regions where these terms are positive are damping. 

Examining further, the second term in brackets and the term on the second line are due to $\partial\fb/\partial v$, which is always damping for the slowing down function. Both of these terms are negligible for $\lres \neq 0$, $\omeganorm < 1$ and $\dl < 1$, which is the case considered here. The first term in brackets is the fast ion drive/damping due to anisotropy $(\partial\fb/\partial\lambda)$, which usually dominates the $\partial\fb/\partial v$ terms except in a very narrow region where $\lambda \approx \linj$. Considering only fast ions with $\vpres > 0$, modes driven by the $\lres = -1$ resonance are destabilized by resonant particles with $\partial\fb/\partial\lambda < 0$ (equivalent to $\lambda > \lambda_0$ for our model distribution), whereas those interacting via the $\lres = 1$ resonance are driven by $\partial\fb/\partial\lambda > 0$ ($\lambda < \lambda_0$). This leads to a useful corollary to this expression without any further simplification: when $1 - \vpres^2/v_0^2 \leq \linj\omegacires$, the integrand does not change sign over the region of integration. As a corollary, 

\begin{equation}
\label{eq:corollary}
1 - \vpres^2/v_0^2 \leq \linj\omegacires \rightarrow 
\left\{
\begin{array}{ll}
\gamma < 0 & \lres = -1 \\ 
\gamma > 0 & \lres = 1 
\end{array}
\right.
\end{equation}

For the single beam distribution in \eqref{eq:Fdistr}, if $1 - \vpres^2/v_0^2 \leq \linj\omegacires$, then modes driven by the $\lres = -1$ resonance (co-propagating) will be strictly damped by fast ions, while those driven by $\lres = 1$ (cntr-propagating) will exclusively be driven by fast ions. This represents a simple sufficient condition for net fast ion drive or damping when this relation between the mode properties ($\krat$ and $\omeganorm$, which determine $\vpres$ through the resonance condition) and fast ion distribution parameters ($v_0$ and $\linj$) is satisfied. 

Moreover, this condition reveals an instability regime unique to slowing down distributions generated by NBI with finite injection energy. This regime was not addressed in the initial studies, which considered either mono-energetic\cite{Belikov1968ZHETF} or bi-Maxwellian\cite{Timofeevv5} distributions for beam ions. Previous studies related to NBI-driven CAEs/GAEs\cite{Gorelenkov2003NF,Kolesnichenko2006POP} also overlooked this regime by implicitly assuming $\vpres \ll v_0$. Consequently, their results were used to interpret experimental observations in NSTX(-U)\cite{Fredrickson2004POP,Fredrickson2017PRL,Fredrickson2018NF} and DIII-D\cite{Heidbrink2006NF} in cases where they may not have been valid. In contrast, this new instability regime can more consistently explain the excitation and suppression of cntr-GAEs observed in NSTX-U,\cite{Fredrickson2017PRL,Belova2019POP} and also suggests that the properties of high frequency modes previously identified as CAEs in DIII-D\cite{Heidbrink2006NF} would in fact be more consistent with those of GAEs. 

Lastly, it is clear from the derivation and discussion in this section that $\lres = \pm 1$ instabilities can occur for any value of $\kperp\rhob$, depending on the parameters of the distribution $(\linj, \vinj)$ and the given mode properties $(\omeganorm,\krat)$. In contrast, in the previously studied regime where $\vpres \ll v_0$ and $\dl \ll 1$, net fast ion drive only occurs for specific ranges of $\kperp\rhob$ when $\omeganorm \ll 1$.\cite{Gorelenkov2003NF} For further understanding of the relationships between the relevant parameters required for instability, analytic approximations or numerical methods must be employed. 

\begin{figure*}[tb]
\hspace*{-2ex}
\subfloat[]{\includegraphics[width = 0.365\textwidth]{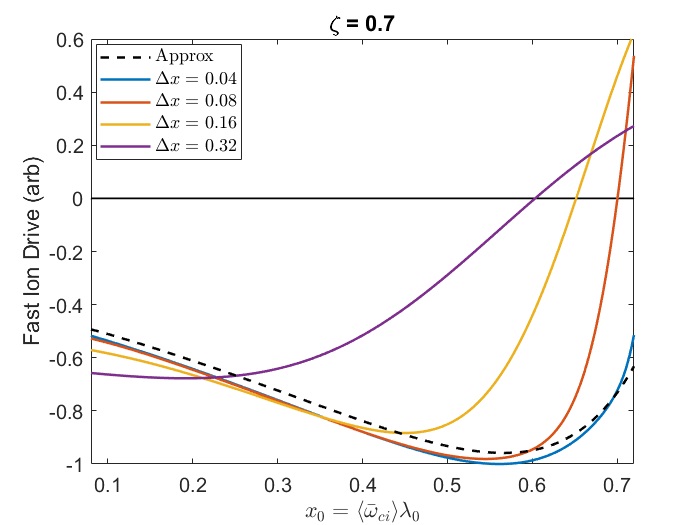}}
\hspace*{-4ex}
\subfloat[]{\includegraphics[width = 0.365\textwidth]{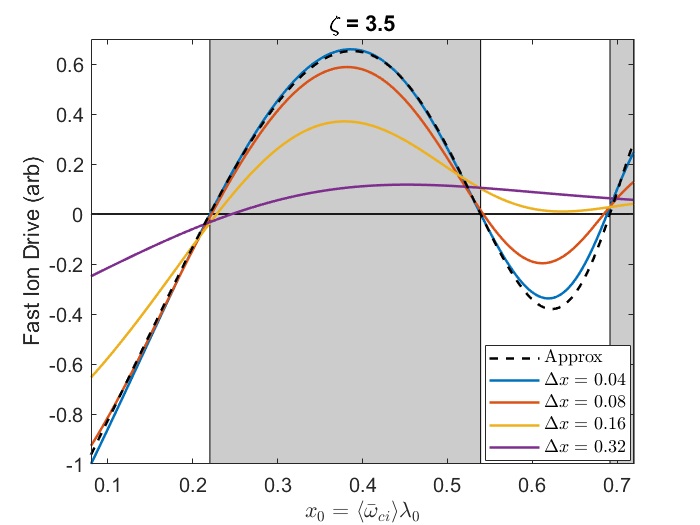}}
\hspace*{-4ex}
\subfloat[]{\includegraphics[width = 0.365\textwidth]{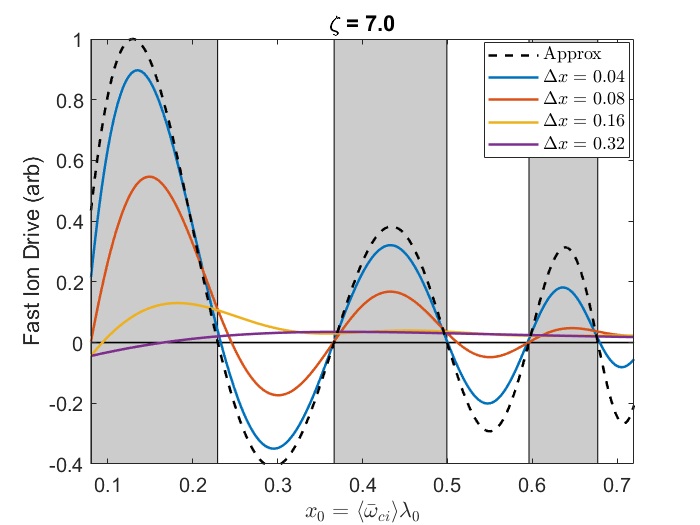}}
\caption{Comparison of numerically integrated growth rate to narrow beam approximation for cntr-GAE with $\bres = 0.2$ as a function of the central trapping parameter of the beam distribution. Black dashed line shows the analytic approximation made in \eqref{eq:gammanarrow} for $\dx = 0.04$ and (a) $\zp = 0.7$, (b) $\zp = 3.5$, and (c) $\zp = 7.0$. Colored curves show numerical integration of \eqref{eq:gammasimp} for different values of $\dx$: blue $\dx = 0.04$, orange $\dx = 0.08$, gold $\dx = 0.16$, and purple $\dx = 0.32$. Shaded regions correspond to regions of drive according to the narrow beam approximation.}
\label{fig:narrowcomp}
\end{figure*}

\section{Approximate Stability Criteria} 
\label{sec:approxstab}

The expression derived in \eqref{eq:gammabeam} can not be integrated analytically, and has complicated parametric dependencies on properties of the specific mode of interest: GAE vs CAE, $\krat$, $\omeganorm$, and the cyclotron coefficient $\lres$ as well as on properties of the fast ion distribution: $\vinj$, $\linj$, and $\dl$. For chosen values of these parameters, the net fast ion drive can be rapidly calculated via numerical integration. Whenever $1 - \vpres^2/v_0^2 \leq \linj\omegacires$, \eqref{eq:corollary} provides the sign of the drive/damping. When this inequality is not satisfied, there are also regimes where approximations can be made in order to gain insight into the stability properties analytically: one where the fast ion distribution is very narrow ($\dl \lesssim 0.10$) and one where it is moderately large $(\dl \gtrsim 0.20$). The former allows comparison with previous calculations,\cite{Gorelenkov2003NF,Kolesnichenko2006POP} while the latter includes the experimental regime where the distribution width in NSTX is typically $\dl \approx 0.30$. In this section, marginal stability criteria will be derived in these regimes. 

\subsection{Approximation of Very Narrow Beam}
\label{sec:narrow}

For the first regime, consider the approximation of a very narrow beam in velocity space. The purpose of this section is to determine when such an approximation can correctly capture the sign of the growth rate. For simplicity, also consider $\omeganorm \ll 1$ so that the anisotropy term dominates and also $\lres/\omegabar \gg x$. Then \eqref{eq:gammabeam} can be re-written as 

\begin{align}
\label{eq:gammasimp}
\frac{\gamma}{\omegaci} &\propto \int_0^{1-\bres} h(x)(x-\xinj)e^{-(x-\xinj)^2/\dx^2}dx \\ 
\text{where } h(x) &= -\frac{\lres C_f}{\dx^2} \frac{x}{(1-x)^2}\frac{\Jlm(\flr(x,\zp))}{1 + \frac{v_0^3}{v_c^3}\left(\frac{\bres}{1-x}\right)^{3/2}}
\end{align} 

If $\dx$ is very small, then the integral is dominated by a contribution in a narrow region $\xinj - \delta < x < \xinj + \delta$ where $\delta \approx 2\dx$. In this region, $h(x)$ can be approximated as a linear function, $h(x) \approx h(\xinj) +  (x-\xinj)h'(\xinj) \plusord{\dx^2}$. So long as $0 < \xinj - \delta$ and $\xinj + \delta < 1 - \bres$, this approximation can be applied:

\begin{align}
\frac{\gamma}{\omegaci} &\appropto h'(\xinj) \int_{\xinj-\delta}^{\xinj+\delta} (x - \xinj)^2e^{-(x-\xinj)^2/\dx^2}dx 
\label{eq:gammanarrow}
\end{align}

The integral is positive, so the sign of the growth rate is equal to the sign of $h'(\xinj)$. Note that this is the same instability regime as studied in previous papers on sub-cyclotron mode stability.\cite{Gorelenkov2003NF,Kolesnichenko2006POP} A comparison of the approximate narrow beam stability criteria to the unapproximated expression for cntr-GAEs with $\bres = 0.2$ is shown in \figref{fig:narrowcomp}. There, the dashed line shows the approximate analytic result \eqref{eq:gammanarrow} plotted as a function of $\xinj$ for $\dx = 0.04$ and different values of $\zp$. Values of $\xinj$ where $h'(\xinj) > 0$ indicate regions where the fast ions are net driving according to this assumption (shaded regions). For comparison, the full expression \eqref{eq:gammasimp} is integrated numerically for each value of $\xinj$ for varying $\dx = 0.04, 0.08, 0.16, 0.32$. This figure demonstrates where the narrow beam approximation correctly determines the sign of the fast ion drive, and how it depends on $\zp$. The curves for $\dx = 0.04$ and $\dx = 0.08$ have essentially the same roots as the analytic expression, whereas the zeros of $\dx = 0.16$ and $\dx = 0.32$ begin to drift away from the approximation or miss regions of instability entirely. The differences are most pronounced for larger values of $\zp$, since this causes the integrand to oscillate more rapidly. Hence, the approximate criteria in \eqref{eq:gammanarrow} is only reliable for $\dx \lesssim 0.10$, especially when $\zp \gg 1$, which is much more narrow than experimental fast ion distributions due to neutral beam injection which have $\dx \approx 0.30$ in NSTX. 

It is unsurprising that this type of approximation fails for realistically large values of $\dx$ since the width of the Gaussian spans nearly the entire integration region. Even for smaller $\dx$, the conclusion from \eqref{eq:gammanarrow} is restricted to situations when both $0 < \xinj - \delta$ and $\xinj + \delta < 1-\bres$ are satisfied. For instance, when $\bres = 0.2$ and $\dx = 0.1$, this expression is only strictly valid for $0.2 < \xinj < 0.6$. 

\subsection{Approximation of Realistically Wide Beam}
\label{sec:wide}

When the beam distribution instead has a non-negligible width in the trapping parameter $\lambda$, a complementary approach can be taken. For $\dx$ sufficiently large, one may approximate $d \exp(-(x-\xinj)^2/\dx^2)/dx \approx -2(x-\xinj)/\dx^2$. This is reasonable for $\xinj - \dx/\sqrt{2} < x < \xinj + \dx/\sqrt{2}$ since this linear approximation is accurate up to the local extrema in this function. When $\dx$ is large, this approximation region may cover nearly the entire region of integration. Throughout this section, $v_c = v_0/2$ will be taken as a representative figure, and the slowing down part of the distribution will be approximated as constant since it makes a small quantitative difference. Then \eqref{eq:gammabeam} may be well-approximated by

\begin{equation}
\gamma \appropto -\int_0^{1-\bres}\frac{x}{(1-x)^2}\Jlm(\flr)\left(\frac{\lres}{\omegabar} - x\right)\left(x - \xinj\right)dx
\label{eq:gammawide}
\end{equation}

This is still not possible to integrate directly because of the Bessel functions with complicated arguments in $\Jlm(\flr)$ since $\flr = \zp\sqrt{x/(1-x)}$. Substituting the values of $\omeganorm$ and $\krat$ from the most unstable modes in \HYM simulations into \eqref{eq:zp} shows that the majority of these modes have $\zp \approx 0.5 \tto 1$, with the largest values being $\zp \approx 3$. Since this parameter controls how rapidly $\Jlm(\flr)$ oscillates, we are motivated to consider two cases separately: the small ($\zp \ll 1$) and large ($\zp \gg 1$) FLR regimes. 

\subsubsection{Small FLR regime \texorpdfstring{$(\zp \ll 1)$}{}}
\label{sec:slow}

For small $\zp$, the argument of the Bessel function will be small for most of the domain. For instance, $x = 1/(1 + \zp^2/\flr^2)$, so when $\zp = 0.5$, the small argument condition $\flr \ll 1$ is true for $x \ll 0.8$, which is the majority of the domain for $\bres$ not too small. 
The leading order approximation to $\Jlm(\flr)$ for $\lres = \pm 1$ and $\flr \ll 1$ is $c \plusord{\flr^2}$ with $c$ constant. For demonstration purposes, it will also be assumed that $\omegabar \ll 1$. This small correction is addressed in \appref{app:omegabar}. With this approximation, \eqref{eq:gammawide} can be simplified and then integrated exactly as 

\begin{align}
\gamma &\appropto -\lres\int_0^{1-\Bres}\frac{x(x-\xinj)}{(1-x)^2}dx
\end{align}

Solving for the marginal stability condition $\gamma = 0$ yields 

\begin{align}
\label{eq:wideslowGAEgamraw}
\xinj &= \frac{1 - \bres^2 + 2\bres\log\bres}{1 - \bres + \bres\log\bres} 
\approx 1 - \bres^{2/3} \\ 
\Rightarrow v_0 &= \frac{\vpres}{\left(1 - \xinj\right)^{3/4}}
\label{eq:wideslowGAEgam}
\end{align}

The serendipitous approximation is better than $1\%$ accurate everywhere. It is arrived at by noticing that \eqref{eq:wideslowGAEgamraw} is a smooth, convex, monotonically decreasing function on $(0,1) \rightarrow (0,1)$, which suggests an \emph{ansatz} of the form $f(x) = 1 - x^p$ for $0 < p < 1$. The choice of $p = 2/3$ is made in order to match the value of the derivative at the $x = 1$ boundary. Note that this stability condition depends implicitly on the mode parameters $\omeganorm$ and $\krat$ through the dependence of $\vpres$, as in \eqref{eq:vpres}. The cases of $\lres = \pm 1$ have the same stability boundary, with an overall sign difference.  Hence, when $\zp \ll 1$, the cntr-propagating $\lres = +1$ CAEs/GAEs are destabilized by fast ion distributions with $v_0 < \vpres/(1 - \xinj)^{3/4}$ and the co-propagating $\lres = -1$ CAEs/GAEs have net fast ion drive when $v_0 > \vpres/(1 - \xinj)^{3/4}$. 

\begin{figure}[tb]
\includegraphics[width = \halfwidth]{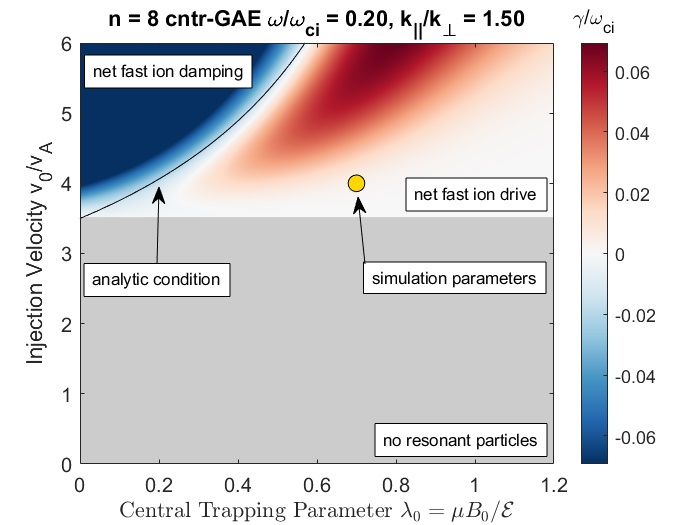}
\caption{Numerical integration of full growth rate expression \eqref{eq:gammabeam} as a function of fast ion distribution parameters $\vinj$ and $\linj$ with $\dx = 0.30$ for a cntr-GAE with properties inferred from \HYM simulations: $\omeganorm = 0.20$ and $\krat = 1.50$, implying $\zp = 0.47$. Red indicates net fast ion drive, blue indicates net fast ion damping, and gray indicates beam parameters with insufficient energy to satisfy the resonance condition. Black curve shows approximate stability condition derived in \eqref{eq:wideslowGAEgam}.}
\label{fig:wideslowGAEfig}
\end{figure} 

It is prudent to compare this approximate analytic condition against the numerical evaluation of \eqref{eq:gammabeam} for a characteristic mode. This is done in \figref{fig:wideslowGAEfig}, where the full expression for fast ion drive of $\lres = +1$ GAE is integrated numerically for a beam distribution with $\dl = 0.30$ (estimated experimental value) and a range of values of $\linj$ and $\vinj$. A representative $n = 8$ cntr-GAE is chosen from \HYM simulations which had $\omeganorm = 0.20$ and $\krat = 1.50$, implying a value of $\zp = 0.47$. The color indicates the sign of the growth rate: red is positive (net fast ion drive), blue is negative (net fast ion damping), while gray is used for beam parameters with insufficient energy to satisfy the resonance condition. The analytic instability condition derived in \eqref{eq:wideslowGAEgam} is shown as the black curve, demonstrating a remarkably good approximation to the full numerical calculation.

Similarly good agreement between the approximation and numerical calculation shown in \figref{fig:wideslowGAEfig} holds even up to $\zp \lesssim 2$ since $\flr = \zp\sqrt{x/(1-x)} \lesssim 1$ is typically still obeyed for most of the integration region in that case, so long as $\bres$ is not too small. Since $\zp = \abs{\omegabar - \lres}/\alpha$ (\eqref{eq:zp}), typically values of $\alpha \gtrsim 0.5$ lead to validity of this regime. When $\zp$ becomes too large, the lowest order Bessel function expansion of $\Jlm(\flr)$ employed in this section is no longer valid over enough of the integration domain for the result to be accurate. For values of $\omegabar$ and $\alpha$ which lead to $\zp \gg 2$, the asymptotic form of the Bessel functions must be used instead to find different stability boundaries, which are derived in the next section. The ``wide beam'' approximate stability conditions remain a good approximation to the numerical calculation for about $0.20 < \dx < 0.80$. If $\dx$ is smaller than this minimum value, the wide beam approximation begins to break down, while $\dx$ larger than the maximum value is where the damping due to the neglected $\partial\fb/\partial v$ term begins to become more important and lead to a nontrivial correction.  

\subsubsection{Large FLR regime \texorpdfstring{$(\zp \gg 1)$}{}}
\label{sec:fast}

Another limit can be explored, that is of the wide beam and rapidly oscillating integrand regime, namely $\zp \gg 1$. This limit is applicable when very large FLR effects dominate most of the region of integration. Based on the most unstable modes found in the \HYM simulations, this is not the most common regime for NSTX-like plasmas, but it can occur and is treated for completeness and comparison to the slowly oscillating results. 

This approximation allows the use of the asymptotic form of the Bessel functions: $J_n(\flr) \like \sqrt{2/\pi \flr}\cos\left(\flr - (2n+1)\pi/4\right) \plusord{\flr^{-3/2}}$, which is very accurate for $\flr > 2$. Note also that $\zp \gg 1$ implies $\alpha \ll 1$ since $\zp = \abs{\lres - \omegabar}/\alpha < 2/\alpha$ for $\abs{\lres} \leq 1 $. 
Since $\alpha \ll 1$, the FLR functions for $\lres = \pm 1$ are well-approximated by $\J{\pm1}{G} \like J_1^2(\flr)/\flr^2 \like 
(1 - \sin(2\flr))/ \flr^3$ for GAEs and $\J{\pm 1}{C}(\flr) \like J_0^2(\flr) \like (1 - \sin(2\flr))/ \flr$ for CAEs. Considering first the case of the $\lres = \pm 1$ GAEs, the relevant integral is 

\begin{align}
\gamma &\appropto -\lres\int_0^{1-\bres} \frac{dx}{\sqrt{x(1-x)}}\left[1 - \sin\left(2\zp\sqrt{\frac{x}{1-x}}\right)\right](x-\xinj) \label{eq:boappx}\\ 
&= -\lres\int_0^{1-\bres} \frac{(x-\xinj)}{\sqrt{x(1-x)}}dx \\ 
&= -\sqrt{\bres(1-\bres)} + (1 - 2\xinj)\arccos\sqrt{\bres}
\label{eq:highpsiGAE}
\end{align}

The first line is \eqref{eq:gammawide} using the asymptotic expansion of the Bessel functions, then the second line is obtained using the stationary phase approximation for rapidly oscillating integrands.\cite{BenderOrszagStationaryPhase} Specifically, the Riemann-Lebesgue lemma\cite{BenderOrszagStationaryPhase} guarantees that $\int_a^b f(t) e^{ixt}dt \rightarrow 0$ for $x\rightarrow \infty$ with integrable $\abs{f(t)}$, which is clear with the substitution of $t = 2\sqrt{x/(1-x)}$ in \eqref{eq:boappx}. Then as before, the marginal stability condition can be found and inverted after an approximation procedure: 

\begin{align}
\xinj &= \frac{1}{2}\left(1 - \frac{\sqrt{\bres(1-\bres)}}{\arccos\sqrt{\bres}}\right) \approx \frac{1}{2}\left(1 - \bres^{2/3}\right) \\ 
\Rightarrow v_0 &= \frac{\vpres}{\left(1 - 2\xinj\right)^{3/4}}
\label{eq:highpsiGAEbound}
\end{align}

The approximation above is found with the same procedure as described for \eqref{eq:wideslowGAEgamraw}, and has a maximum relative error of $3\%$. Interestingly, this condition is similar to the one derived for $\zp \ll 1$ except that $(1 - \xinj)$ has been replaced by $(1 - 2\xinj)$. This condition describes the boundary for $\lres = \pm 1$ GAEs, with $v_0 > \vpres/(1 - 2\xinj)^{3/4}$ indicating net fast ion drive for $\lres = -1$ co-GAEs and net fast ion damping for $\lres = +1$ cntr-GAEs. 

\begin{figure}[tb]
\includegraphics[width = \halfwidth]{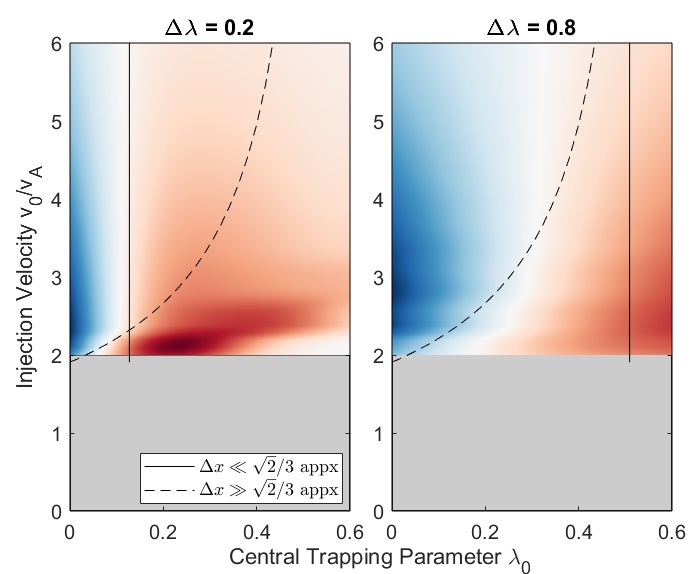}
\caption{Comparison of approximations for marginal fast ion drive for cntr-GAEs with $\zp \gg 1$ and $\dx \lesssim \sqrt{2}/3$ vs $\dx \gtrsim \sqrt{2}/3$. Left is the former (with $\dx = 0.20$) and right is the latter (with $\dx = 0.80$). Both use $\omeganorm = 0.3$ and $\krat = 0.07$ so that $\zp = 8.6$, and also $\omegacires = 0.9$. Red indicates net fast ion drive, while blue indicates net fast ion damping, and gray indicates beam parameters with insufficient energy to satisfy the resonance condition. The vertical line is the approximate marginal stability boundary of $\xinj = \dx/\sqrt{2}$, valid when $\dx \lesssim \sqrt{2}/3$ for $\zp \gg 1$. The dashed curve is the approximate marginal stability boundary of $\vinj = \vpres/(1 - 2\linj\omegacires)^{3/4}$, valid when $\dx \gtrsim \sqrt{2}/3$ for $\zp \gg 1$.}
\label{fig:widefastGAEfig}
\end{figure} 

\newcommand{\vroom}{\vphantom{{\Huge text}}}
\newcommand{\hroom}{\hspace{1ex}}
\newcommand{\chead}[1]{\multicolumn{1}{c}{#1}}
\begin{table*}[t]\centering
\renewcommand\arraystretch{1.5}
\begin{tabular}{cc}
\begin{tabular}[t]{l l l l}
\multicolumn{4}{c}{GAE fast ion drive conditions} \\ \hline\hline
 & & \chead{$\lres=+1$ (cntr)} & \chead{$\lres=-1$ (co)} \\ 
\hline
$\zp \lesssim 2$ & & $v_0 < \dfrac{\vpres}{(1 - \xinj)^{3/4}}$ \hroom & $v_0 > \dfrac{\vpres}{(1 - \xinj)^{3/4}}$ \vroom \\
\multirow{2}{3em}{$\zp \gg 2$} & $\dx \lesssim \sqrt{2}/3$ \hroom & $\xinj > \dx/\sqrt{2}$ & $\xinj < \dx/\sqrt{2}$ \vroom \\
& $\dx \gtrsim \sqrt{2}/3$ & $v_0 < \dfrac{\vpres}{(1 - 2\xinj)^{3/4}}$ & $v_0 > \dfrac{\vpres}{(1 - 2\xinj)^{3/4}}$ \vroom \vspace{1ex} \\ 
\hline\hline
\end{tabular}
\quad
%
\begin{tabular}[t]{l l l}
\multicolumn{3}{c}{CAE fast ion drive conditions} \\ \hline\hline
 & \chead{$\lres = +1$ (cntr)} 
 & \chead{$\lres=-1$ (co)} \\ 
\hline
$\zp \lesssim 2$ \hroom\hroom & $v_0 < \dfrac{\vpres}{(1 - \xinj)^{3/4}}$ \hroom\hroom & $v_0 > \dfrac{\vpres}{(1 - \xinj)^{3/4}}$ \hroom\vroom \\
$\zp \gg 2$ & $v_0 < \dfrac{\vpres}{(1 - \xinj)^{5/6}}$ & $v_0 > \dfrac{\vpres}{(1 - \xinj)^{5/6}}$ \vroom \vspace{1ex}\\ 
\hline\hline
\end{tabular}
\end{tabular}

\caption{Approximate net fast ion drive conditions for GAEs and CAEs driven by $\lres = \pm 1$ resonances in the wide beam approximation, valid for $0.2 < \dx < 0.8$. The quantity $\zp = \kperp\vpres/\omegaci$ is the ``modulation parameter" (see \eqref{eq:zp}) and $x_0 = \lambda_0\omegacires = \vperpz^2/v_0^2$.}
\label{tab:appxcons}
\end{table*}

When compared to the exact numerical calculation in this regime, \eqref{eq:highpsiGAEbound} captures the qualitative feature that the stability boundary occurs at much lower $\xinj$ than in the low $\zp$ regime. However, the quantitative agreement is not as good unless $\dx \approx 0.6$. For smaller values of $\dx$, the approximations become poor for large $x \gtrsim \xinj + \dx \sqrt{2}$ where the Gaussian decay would tend to dominate the diverging term $1/\sqrt{1-x}$ at $x\rightarrow 1$. This can be seen in \figref{fig:widefastGAEfig} where the marginal stability boundary approaches a vertical asymptote. To capture this behavior, the wide beam approximation can still be used, but with the integration running from $x = 0 \tto a = \xinj + \dx\sqrt{2}$ instead of $x = 0 \tto 1 - \bres$ to replicate the decay expected beyond this region. Then, the fast ion drive is approximately 

\begin{align}
%
\gamma &\appropto \lres\int_0^{a} \frac{(x-\xinj)}{\sqrt{x(1-x)}}dx \\ 
&= -\sqrt{a(1 - a)} + (1 - 2\xinj)\arcsin\sqrt{a} \\ 
\Rightarrow \xinj &= \frac{1}{2}\left[1 - \frac{\sqrt{a(1-a)}}{\arcsin\sqrt{a}}\right] \\ 
&\approx \frac{1}{2}\left[1 - (1 - \xinj - \dx\sqrt{2})^{2/3}\right] 
\label{eq:highpsiGAElim}
\end{align}

The approximation in the last line has a maximum global error of $3\%$. If $\xinj + \dx\sqrt{2}$ is close to 1, then the term in round braces is small, and the limit of $\xinj \rightarrow 1/2$ is recovered from \eqref{eq:highpsiGAEbound}. Hence, the other case of interest is when $\xinj + \dx\sqrt{2}$ is small, in which case a linear approximation admits a solution for \eqref{eq:highpsiGAElim} of $\xinj = \dx/\sqrt{2}$, which gives much better agreement with the numerically calculated boundary shown in \figref{fig:widefastGAEfig}. Hence, \eqref{eq:highpsiGAEbound} is applicable for $\dx \gtrsim \sqrt{2}/3$, whereas $\xinj = \dx/\sqrt{2}$ gives the limiting boundary for smaller $\dx$. 

A similar procedure can be used to approximate the marginal stability boundaries for CAEs, however it is rare for CAEs to be excited with $\zp \gg 1$ for the parameters studied here. This is because the CAE dispersion combined with the resonance condition yields $\zp \approx \omegabar\vpres/\va$ for $\zp \gg 1$, which can not be very large for $\vinj < 6$ considering $\vpres \like v_0/2$ is common, as is $\omeganorm \like 1/2$. The case is different for GAEs since their dispersion yields a parallel resonant velocity that is independent of $\alpha$, such that $\zp$ can be made arbitrarily large by choosing $\alpha$ sufficiently small without constraining the size of $\vpres/\va$. The case of $\zp \gg 1$ for CAEs with $\lres = \pm 1$ is treated in \appref{app:fastcae}. 

\subsection{Summary of Necessary Conditions for Net Fast Ion Drive}

For clarity, it is worthwhile to summarize all of the conditions for net fast ion drive derived in this section and remind the reader of their respective ranges of validity. When $1 - \vpres^2/v_0^2 \leq \linj\omegacires$ is satisfied, $\lres = -1$ modes will be net damped by fast ions, while those interacting via the $\lres = 1$ resonance will be net driven. All other results address the scenarios when this inequality is not satisfied, which is the parameter regime considered by previous authors.\cite{Gorelenkov2003NF,Kolesnichenko2006POP} When $\dl$ is sufficiently small $(\dl \lesssim 0.10)$, the narrow beam approximation can be made, which yields \eqref{eq:gammanarrow} and implies that net drive vs damping depends on the sign of $h'(\xinj)$. When $\dl$ is sufficiently large $(0.20 \lesssim \dl \lesssim 0.80)$, the wide beam approximation is justified. This includes the nominal NSTX case of $\dl \approx 0.3$. For most of the unstable modes in \HYM simulations, $\zp \lesssim 2$ is also valid, which facilitates the results obtained in the case of a wide beam with small FLR effects. The complementary limit of $\zp \gg 2$ is also tractable when the beam is sufficiently wide, though this is not the typical case in NSTX conditions, except for some low $n$ cntr-GAEs. All conditions for the cases involving wide beams are organized in \tabref{tab:appxcons}. 

\section{Preferential excitation as a function of mode parameters} 
\label{sec:stabao}

For fixed beam parameters, the theory can determine which parts of the spectrum may be excited -- complementary to the previous figures which addressed how the excitation conditions depend on the two beam parameters for given mode properties. Such an examination can also illustrate the importance of coupling between the compressional and shear branches due to finite frequency effects on the most unstable parts of the spectra. All fast ion distributions in this section will be assumed to have $\dl = 0.3$ and $\omegacires = 0.9$ for the resonant ions. 

\begin{figure*}[tb]
\subfloat[]{\includegraphics[width = \halfwidth]{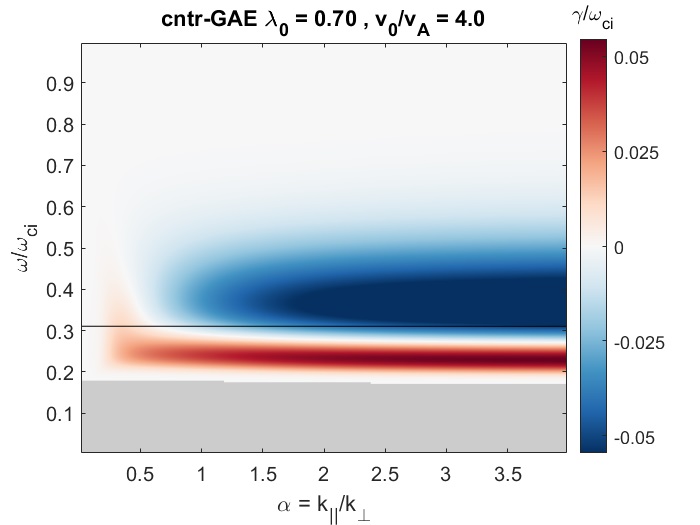} \label{fig:GAEao-cntr}}
\subfloat[]{\includegraphics[width = \halfwidth]{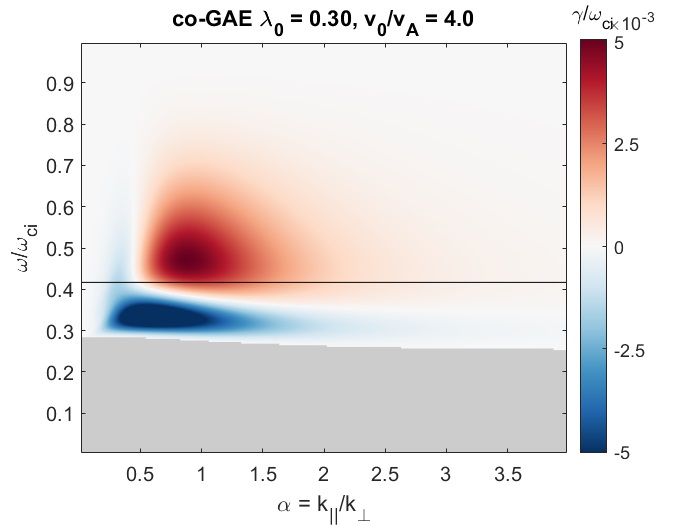} \label{fig:GAEao-co}}
\caption{Numerically calculated fast ion drive/damping for GAEs as a function of $\omegabar = \omeganorm$ and $\alpha = \krat$, when driven by a beam distribution with (a) $\linj = 0.7$ for cntr-GAEs and (b) $\linj = 0.3$ for co-GAEs. Also, $\vinj = 4.0$, $\dl = 0.3$, and assuming $\omegacires \approx 0.9$. Red corresponds to net fast ion drive, blue to damping, and gray to regions excluded by the resonance condition. Black line is the marginal frequency for fast ion drive predicted by the approximate analytic conditions in \eqref{eq:GAEmfreqrange} and \eqref{eq:GAEpfreqrange}.}
\label{fig:GAEao}
\end{figure*} 

\subsection{GAE Stability}
\label{sec:GAEao}

Consider first the GAEs. As a consequence of the approximate dispersion $\omega \approx \abs{\kpar}\va$, the necessary condition $\vpres < v_0$ for resonant interaction, and the net fast ion drive condition derived in \eqref{eq:wideslowGAEgam},
the region in $(\omegabar,\alpha)$ space corresponding to net fast ion drive in the typical case of $\zp\lesssim 1$ is nearly independent of $\alpha$. For counter-propagating modes with $\lres = 1$,   

\begin{align}
\frac{\omegacires}{\vinj + 1} < 
\left(\frac{\omega}{\omegaci}\right)_{\lres = 1}^{GAE} < 
\frac{\omegacires}{\vinj\left(1 - \linj\omegacires\right)^{3/4} + 1}
\label{eq:GAEmfreqrange}
\end{align}

Hence, the theory predicts a relatively small band of unstable frequencies. Larger $\vinj$ decreases both boundaries, leading to a range of unstable frequencies of about $(\omega_\text{max} - \omega_\text{min})/\omegaci \approx 10 - 20\%$. 

For co-propagating GAEs driven by $\lres = -1$, there is instead a lower bound on the unstable frequencies: 

\begin{align}
\frac{\omegacires}{\vinj(1-\linj\omegacires)^{3/4}-1} < \left(\frac{\omega}{\omegaci}\right)_{\lres = -1}^{GAE} < 1
\label{eq:GAEpfreqrange}
\end{align}

These conditions can be compared against the net fast ion drive calculated from \eqref{eq:gammabeam} as a function of $\omeganorm$ and $\krat$ for a distribution with $\vinj = 4$. Central trapping parameters $\linj = 0.7$ and $\linj = 0.3$ are used for cntr- and co-GAEs, respectively. The calculation is shown in \figref{fig:GAEao}. The simple analytic conditions are reasonably close to the true marginal stability on these figures. Further improved agreement could be achieved by substituting the full coupled dispersions from \eqref{eq:stixdisp} into the formula for $\vpres$ in \eqref{eq:vpres}, though the resulting boundaries would be implicit. The deviation from the analytic line on the figure at very low $\alpha$ is due to the inapplicability of the assumption $\zp \ll 1$ which was used to derive the approximate boundary, since very low $\alpha$ implies very large $\zp$ according to \eqref{eq:zp}, which has a different instability condition, as discussed in \secref{sec:fast}. 

The variation of the growth rate as a function of $\alpha$ is due to coupling between the shear and compressional branches, as well as FLR effects, contained within \eqref{eq:stixdisp} and \eqref{eq:Jlm}. For large $\alpha \gg 1$, the FLR functions in \eqref{eq:Jlmbigappx-gae} are valid, and as discussed previously, $\alpha \rightarrow \infty$ is equivalent to $\flr \rightarrow 0$. For the cntr-GAEs, $\J{1}{G} \propto J_0^2$, which peaks at $\flr = 0$, thus explaining why the growth rate in \figref{fig:GAEao-cntr} increases monotonically with $\alpha$ for the cntr-GAE, and eventually saturating. In contrast, the co-GAEs have $\J{-1}{G} \propto J_2^2$ in this limit, which vanishes for $\flr \rightarrow 0$. When coupling with the compressional branch is not taken into account, the co-GAE would also have its growth rate strictly increasing with $\alpha$ since it would have the same FLR function as the cntr-GAE. 

Conversely, $\alpha \rightarrow 0$ implies $\flr \rightarrow \infty$, where all Bessel functions of the first kind $J_\lres(\flr)$ decay to zero, such that the net drive vanishes for small $\alpha$. For the co-GAE, the growth rate decreasing at both large and small $\alpha$ results in a local maximum in the growth rate at $\alpha \like 1$. When the coupling is neglected, the maximum co-GAE growth rate is increased by a factor of 4 relative to when coupling is included (in addition to being shifted from $\alpha \like 1$ to $\alpha \rightarrow \infty$), whereas the cntr-GAE growth rate is hardly affected.

\begin{figure*}[tb]
\subfloat[]{\includegraphics[width = \halfwidth]{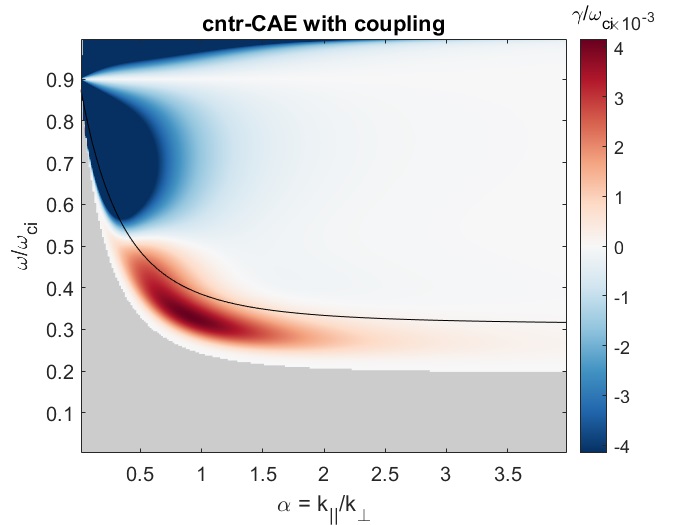}\label{fig:CAEao-full}}
\subfloat[]{\includegraphics[width = \halfwidth]{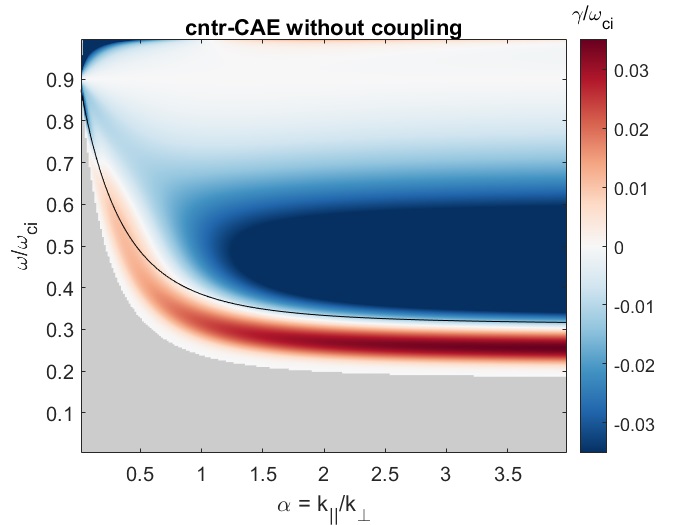}\label{fig:CAEao-simp}}
\caption{Numerically calculated fast ion drive/damping for cntr-CAEs as a function of $\omegabar = \omeganorm$ and $\alpha = \krat$ when coupling to the shear branch is (a) included and (b) neglected. In both calculations, the modes are driven by a beam distribution with $\linj = 0.7$, $\vinj = 4.0$, $\dl = 0.3$, and assuming $\omegacires \approx 0.9$. Red corresponds to net fast ion drive, blue to damping, and gray to regions excluded by the resonance condition. Black line is the marginal frequency for fast ion drive predicted by the approximate analytic condition in \eqref{eq:CAEmfreqrange}.}
\label{fig:CAEao}
\end{figure*} 

\subsection{CAE Stability}
\label{sec:CAEao}

The cntr-CAEs also have a band of unstable frequencies, though this band also depends on $\alpha$. The analogous inequalities using the approximate $\omega \approx k\va$ are

\begin{align}
\frac{\omegacires}{\frac{\abs{\kpar}v_0}{k\va} + 1} < 
\left(\frac{\omega}{\omegaci}\right)_{\lres = 1}^{CAE} < 
\frac{\omegacires}{\frac{\abs{\kpar}v_0}{k\va}\left(1 - \linj\omegacires\right)^{3/4} + 1}
\label{eq:CAEmfreqrange}
\end{align}

The comparison between the full numerical calculation of fast ion drive as a function of $\omegabar,\alpha$ for cntr-CAEs against this approximate boundary is shown in \figref{fig:CAEao}, both when coupling to the shear branch is (a) included and (b) neglected. The agreement between the approximate condition and the numerical marginal stability is quite reasonable in both cases. These two calculations are shown in order to highlight the importance of including this coupling, which comes from finite $\omeganorm$ and FLR effects. Consider first the simpler case when no coupling is present. Then the growth rate increases monotonically with $\alpha$ like it did for the cntr-GAE. The difference between \eqref{eq:CAEmfreqrange} for the cntr-CAEs and \eqref{eq:GAEmfreqrange} for the cntr-GAEs is the additional factor of $\abs{\kpar}/k = \alpha/\sqrt{1 + \alpha^2}$ for the CAEs, which tends to one for large $\alpha$, where \figref{fig:CAEao-simp} and \figref{fig:GAEao-cntr} agree with similar growth rates. 

As was the case with the co-GAEs, the effect of coupling between the two branches is also significant for the cntr-CAEs, and for similar reasons. When coupling is included, \eqref{eq:Jlmbigappx-cae} shows that when $\alpha \gg 1$ for cntr-CAEs, $\J{1}{C} \propto J_2^2$, which goes to zero for small $\flr$. In the approximation of no coupling, instead $\J{1}{C} \propto J_0^2$, which is maximized at $\flr = 0$, just as $\J{1}{G}$ is, explaining the agreement between \figref{fig:CAEao-simp} and \figref{fig:GAEao-cntr} at large $\alpha$. As with the GAEs, the CAE growth rates go to zero for $\alpha \rightarrow 0$ since this is the $\flr \rightarrow \infty$ limit of the Bessel functions, where they decay. Hence, the cntr-CAE has a maximum in its growth rate near $\alpha \like 1$ just as the co-GAE did in the previous section. Likewise, the inclusion of coupling reduces the maximum cntr-CAE growth rate by almost an order of magnitude for the beam parameters used in \figref{fig:CAEao}. It is worth pointing out that the cntr-GAE growth rates are larger than those for the cntr-CAEs at nearly every set of mode and beam parameters, possibly explaining why the GAEs were more frequently observed in NSTX experiments. This may also explain why initial value simulations of NSTX with the \HYM code finds unstable cntr-GAEs but not cntr-CAEs.\cite{Lestz2018APS,Lestz2019sim} 

The analysis of this section shows that coupling between the two branches (due to two-fluid effects in this model) is important in determining the growth rate of the cntr-CAEs and co-GAEs via their influence on the FLR effects from the fast ions. Hence, a two fluid description of the thermal plasma (such as Hall-MHD) may be important in order to accurately model cntr-CAEs and co-GAEs. 

\section{Experimental Comparison} 
\label{sec:expcomp}

\begin{figure}[tb] 
\includegraphics[width = \halfwidth]{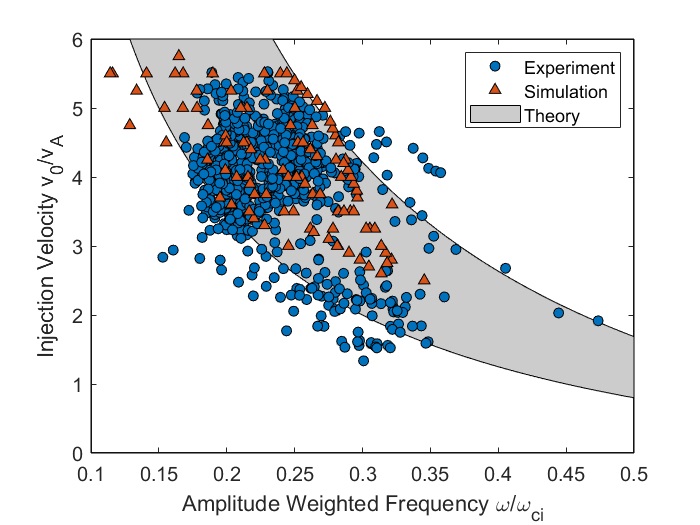}
\caption{Comparison between theory, simulations, and experiment. (a) Blue circles represent amplitude-weighted quantities from 50 ms time windows of NSTX discharges identified as having mostly cntr-GAE activity. Red triangles show cntr-GAEs excited in \HYM simulations. Theory predicts net fast ion drive in the shaded region between the two curves, as in \eqref{eq:GAEmfreqrange}.}
\label{fig:allcomp}
\end{figure}

An experimental database of CAE and GAE activity in NSTX has previously been compiled and analyzed.\cite{Tang2017TTF} This database includes approximately 200 NSTX discharges, separated into over 1000 individual 50 ms analysis windows. For each time slice, fluctuation power-weighted averages of mode quantities were calculated. The simplified instability conditions derived here relating the beam injection parameters to the mode parameters depends only on $\linj,\vinj,\omegabar$ for GAEs, which are relatively well-known and measured quantities. Hence, a comparison can be made between the marginal fast ion drive conditions and the experimental observations, shown in \figref{fig:allcomp}. This comparison assumes that the $\zp \lesssim 1$ regime (which described the most unstable modes in \HYM simulations) is valid for the experimental modes. 

The blue circles are amplitude-weighted observations in discharges with \Alfvenic activity determined to be predominantly GAE-like. Specifically, the selected time slices satisfy $-10 \leq \avg{n} \leq -4$, $\avg{f} > 200$ kHz, $T_e > 500$ eV, and $\Pb > 1$ MW. These properties were found to correlate with GAE-like modes dominating the spectrum from inspection of the database. 

The red triangles represent unstable cntr-GAEs from \HYM simulations with $\linj = 0.5 - 0.9$, covering the typical range for NSTX NBI distributions. The theory developed in this paper predicts net fast ion drive in the shaded region between the two curves. Further analysis of the linear simulation results shown on \figref{fig:allcomp} will be described in detail in a forthcoming paper.\cite{Lestz2019sim} The simulation set up and properties of the modes can be found in \citeref{Lestz2018POP}. The simulations used equilibrium profiles from the well-studied H-mode discharge $\#141398$,\cite{Fredrickson2013POP,Crocker2013NF,Crocker2017NF,Belova2017POP} and fast ion distributions with the same $(\lambda,v)$ dependence studied in this work, and given in \eqref{eq:Fdistr}. The peak fast ion density in all cases is $n_b/n_e = 5.3\%$, matching its experimental value in the model discharge. 

The theoretically predicted unstable region according to \eqref{eq:GAEmfreqrange} lies in the shaded region between the two curves, which was calculated with $\omegacires = 0.9$, motivated by the mean value of the resonant fast ions in \HYM simulations across a wide range of simulation parameters, and also $\linj \approx 0.7$ as a characteristic value of the NSTX beam geometry. There is strong agreement, especially considering the variety of assumptions required to derive the simplified stability boundaries. When evaluating the instability bounds for the specific values of $\linj$, $\vinj$, and $\omeganorm$ for each data point shown in the figure, $82\%$ of the experimental points are calculated to be theoretically unstable, and $94\%$ of the simulation points. 

An analogous comparison would be more difficult to perform for the other modes discussed in this paper. First, co-propagating GAEs have not yet been observed in experiments since their excitation requires much smaller $\linj$ than was possible on NSTX. If they are observed in future NSTX-U experiments, as they could be in low field scenarios with the new, more tangential beam sources, a comparison could be made. Moreover, there appear to be fewer discharges dominated by cntr-CAEs than cntr-GAEs, hence requiring time-intensive inspection of many discharges in order to confidently identify cntr-CAE modes for comparison. The cntr-CAE instability boundaries (given in \eqref{eq:CAEmfreqrange}) also depend on both $\omeganorm$ and $\krat$, increasing the parameter space of the comparison. Nonetheless, these would be interesting avenues for further cross-validation.   

\section{Summary and Discussion}
\label{sec:summary}

The fast ion drive/damping for compressional (CAE) and global (GAE) \Alfven eigenmodes has been investigated analytically for a model slowing down, beam-like fast ion distribution in 2D velocity space, such as distributions generated by neutral beam injection in NSTX. Growth rate expressions previously derived by Gorelenkov\cite{Gorelenkov2003NF} and Kolesnichenko\cite{Kolesnichenko2006POP} were generalized to retain all terms in $\kperp\rhob$, $\krat$ and $\omeganorm$ for sub-cyclotron modes in the local approximation driven by the Doppler-shifted ordinary $(\lres = 1)$ and anomalous $(\lres = -1)$ cyclotron resonances. This general expression for fast ion drive was evaluated numerically to determine the dependence of the fast ion drive/damping on key distribution parameters (injection velocity $\vinj$ and central trapping parameter $\linj = \mu B_0/\W$) and mode parameters (normalized frequency $\omeganorm$ and direction of propagation $\krat$) for each mode type and resonance. Retaining finite $\omeganorm$ and $\krat$, a source of coupling between the shear and compressional branches, was found to be responsible for significantly modifying the cntr-CAE and co-GAE growth rate dependence on $\krat$. 

The derived growth rate led to an immediate corollary: when $1 - \vpres^2/v_0^2 \leq \linj\omegacires$, cntr-propagating modes are strictly driven by fast ions while co-propagating modes are strictly damped. This condition occurs due to a finite beam injection energy, and it uncovers a new instability regime that was not considered in previous studies except recently in \citeref{Belova2019POP}, which were valid only in the $\vpres \ll v_0$ limit. For cases where when $1 - \vpres^2/v_0^2 \leq \linj\omegacires$ is not satisfied, approximate methods were employed to derive conditions necessary for net fast ion drive. Previous analytic conditions were also limited to delta functions in $\lambda$, which are a poor approximation for fast ions generated by NBI. In this work, broad parameter regimes were identified which allow for tractable integration, leading to the first compact net fast ion drive conditions as a function of fast ion and mode parameters which properly integrate over the full beam-like distribution. For the narrow beam case discussed in \secref{sec:narrow}, the sign of the growth rate depends on a function of $\linj$ only, similar to the instability regime studied previously.\cite{Gorelenkov2003NF,Kolesnichenko2006POP} Numerical integration showed that this result was only reliable for beams much more narrow $(\dl \lesssim 0.1)$ than those in experiments $(\dl \approx 0.3)$, underscoring the limitations of past results. In particular, those previous studies identified $\kperp\rhob > 1$ and $\kperp\rhob > 2$ as the most unstable parameters for cntr-CAE and cntr-GAE instabilities, respectively, whereas this work demonstrates that these instabilities may be excited for any value of $\kperp\rhob$, with $\kperp\rhob \lesssim 1$ instabilities perhaps more common for NSTX conditions. 

The approximation of a sufficiently wide beam $(\dl \gtrsim 0.2)$ in conjunction with a small or large FLR assumption allowed the derivation of very simple conditions for net fast ion drive, summarized in \tabref{tab:appxcons}. These expressions depend on the fast ion injection velocity $\vinj$, central trapping parameter $\linj$, and mode properties $\omeganorm$, $\krat$ which determine $\vpres$ along with the cyclotron resonance coefficient $\lres$. It is found that the wide beam, small FLR assumption is valid over a wide enough range of parameters ($\zp = \kperp\vpres/\omegaci \lesssim 2$) that it encompasses the typical conditions for NSTX fast ions and properties of the most unstable CAEs/GAEs inferred from experiments and simulations. 

Comparison between full numerical evaluation of the exact analytic expression and the approximate stability boundaries demonstrate excellent agreement within the ranges of applicability. These regimes include fast ion parameters motivated by \TRANSP/\NUBEAM modeling of NSTX beam profiles, as well as properties $(\omeganorm,\krat)$ of the most unstable modes excited in hybrid simulations with the \HYM code.\cite{Lestz2019sim} In addition to providing insight into an individual mode's growth rate as a function of fast ion parameters, the new instability conditions also yield information about the properties of the unstable modes for a fixed beam distribution. Namely, cntr-propagating GAEs are unstable for a specific range of frequencies (as a function of beam parameters) nearly independent of $\krat$, whereas cntr-CAEs are more sensitive to $\krat$. This condition for cntr-GAEs compares well against NSTX data across many discharges, providing support for the theoretical underpinnings of the growth rate calculation, as well as the series of mathematical approximations made to arrive at these compact marginal stability conditions. 

The approximate conditions for net fast ion drive were only made possible by a series of simplifications, which should be kept in mind when applying these results. Integration over space and $\pphi$ were neglected, restricting the analysis to 2D phase space. Moreover, the derived stability boundaries do not include damping on the background plasma, such that net fast ion drive as calculated in this paper is a necessary but not sufficient condition for overall instability. Including the electron Landau damping rate and the continuum/radiative damping due to interaction with the \Alfven continuum is an area for future work.

The results derived here can be applied in the future to help interpret experimental results and improve physics understanding of first principles simulations. Ideally, they can be used to guide expectations about the spectrum of unstable modes that will be generated by a specific neutral beam configuration. For instance, if a specific mode is driven unstable by an initial beam distribution, these expressions show where additional neutral beam power may be added that would act to stabilize this mode, or drive it further unstable, if desired. This enables systematic analysis and prediction of scenarios like those of the cntr-GAE stabilization observed in NSTX-U.\cite{Fredrickson2017PRL,Fredrickson2018NF,Belova2019POP}  

\section{Acknowledgments}
\label{sec:acknowledgments}

The authors are grateful to E.D. Fredrickson for providing additional experimental data for comparison, as well as A.O. Nelson for fruitful discussions. The simulations reported here were performed with computing resources at the National Energy Research Scientific Computing Center (NERSC). The data required to generate the figures in this paper are archived in the NSTX-U Data Repository ARK at the following address:
\url{http://arks.princeton.edu/ark:/88435/dsp011v53k0334}. This research was supported by the U.S. Department of Energy (NSTX contract DE-AC02-09CH11466 and DE-SC0011810).

\appendix 

\section{Correction for finite frequency in small FLR regime \texorpdfstring{($\zp\ll 1$)}{}}
\label{app:omegabar}

Here, the correction due to finite $\omegabar$ for the wide beam, low $\zp$ approximation is addressed. This term was neglected in \secref{sec:wide}. Including this term, the integral of interest is 

\begin{align}
\gamma &\appropto \int_0^{1-\bres} \frac{(x - \xinj)}{(1-x)^2}\left(\frac{\lres}{\omegabar} - x\right)dx = 0 \\ 
\Rightarrow \xinj &= \frac{\lres f(\bres) + \omegabar g(\bres)/2}{\lres h(\bres) - \omegabar f(\bres)} \\ 
f(\bres) &= 1 - \bres^2 + 2\bres\log\bres \\ 
g(\bres) &= -2 - 3\bres + 6\bres^2 - \bres^3 - 6\bres\log\bres \\ 
h(\bres) &= 1 - \bres + \bres\log\bres 
\end{align}

This function can be approximated to leading order in $\omegabar < 1$, and will take advantage of the known approximation from earlier $f(\bres)/h(\bres) \approx 1 - \bres^{2/3}$. 

\begin{align}
\xinj &= \frac{f(\bres)}{h(\bres)} + \frac{\omegabar}{\lres}\left[\left(\frac{f(\bres)}{h(\bres)}\right)^2 + \frac{g(\bres)}{2h(\bres)}\right] \\ 
&\approx 1-\bres^{2/3} - \frac{\omegabar}{8\lres}\bres^{2/3}\left(1-\bres^{2/3}\right)^2
\end{align}

The second term in the second line is the approximation to the function in brackets. Again using $\omegabar$ as a small parameter, assume a solution of the form $\bres = \bres_0 + \omegabar\bres_1$ where $\bres_0 = (1-\xinj)^{3/2}$. Then the leading order correction in $\omegabar$ to the $\omegabar\rightarrow 0$ solution found in \secref{sec:wide} is 

\begin{align}
v_0 = \frac{\vpres}{\left(1 - \xinj\right)^{3/4}}\left(1 + \frac{3\omegabar\xinj^2}{32\lres}\right)
\end{align}

\section{Large FLR regime for CAEs \texorpdfstring{$(\zp \gg 1)$}{}}
\label{app:fastcae}

Using the large $\zp \gg 1$ (equivalently small $\alpha \ll 1$) expansion for CAEs with $\lres = \pm 1$ gives  $\J{\pm 1}{C}(\flr) \like J_0^2(\flr) \like (1 - \sin(2\flr))/ \flr$. As in \secref{sec:fast}, the rapidly varying $\sin(2\flr)$ will average to zero in the integral, leaving 

\begin{align}
\gamma &\appropto -\lres\int_0^{1-\bres} \frac{\sqrt{x}(x-\xinj)}{(1-x)^{3/2}}dx 
\label{eq:highpsiCAE}
\end{align}

Integrating and finding the marginal stability condition $\gamma = 0$ results in 

\begin{align}
\label{eq:highpsiCAEcrit}
\xinj &= \frac{8\sqrt{\bres^{-1}-1} + 4\sqrt{\bres(1-\bres)} - 3\pi - 6\arctan\left(\frac{1-2\bres}{2\sqrt{\bres(1-\bres)}}\right)}
{8\left(\sqrt{\bres^{-1}-1} - \arccos\sqrt{\bres}\right)} \\ 
&\approx 1 - \bres^{3/5} \Rightarrow v_0 = \frac{\vpres}{(1 - \xinj)^{5/6}}
\label{eq:highpsiCAEcon}
\end{align}

The approximation in \eqref{eq:highpsiCAEcon} has a maximum global error of $3\%$. The instability condition for cntr-propagating modes $(\lres = 1)$ is $v_0 < \vpres/(1 - \xinj)^{5/6}$, while the co-propagating modes $(\lres = -1)$ are driven for $v_0 > \vpres/(1 - \xinj)^{5/6}$. 
%
\bibliography{all_bib}
\end{document}

%% file: header.tex
\usepackage{color}

\usepackage[caption=false]{subfig}
\usepackage{graphicx,booktabs,array}
\usepackage{multirow}
\DeclareGraphicsExtensions{.pdf,.png,.jpg,.eps}
\usepackage{epsfig}
\usepackage{epstopdf}
\usepackage{amssymb}
\usepackage{amsmath}
\usepackage{amsfonts}
\usepackage{mathrsfs}
\usepackage{amsthm}
\usepackage{float}
\usepackage{amsbsy}
\usepackage{bm}
\usepackage{grffile}
\usepackage{hyperref}
\hypersetup{
    colorlinks = true,
    citecolor = blue,
    urlcolor = blue,
    linkcolor = blue,
}
\usepackage{courier}
\usepackage{upgreek}
\usepackage{xspace}
\usepackage{xcolor}

\newcommand{\ie}{\textit{i}.\textit{e}. }
\newcommand{\eg}{\textit{e}.\textit{g}. }

\newcommand{\Alfven}{Alfv\'{e}n }
\newcommand{\Alfvenic}{Alfv\'{e}nic }

\newcommand{\va}{v_A}

\newcommand{\vb}{v_0}

\newcommand{\omegaci}{\omega_{ci}}
\newcommand{\omegacio}{\omega_{ci0}} 

\newcommand{\omeganorm}{\omega / \omegaci}

\newcommand{\omegabar}{\bar{\omega}}
\newcommand{\omegace}{\omega_{ce}}
\newcommand{\omegape}{\omega_{pe}}

\newcommand{\omegaps}{\omega_{ps}}
\newcommand{\omegacs}{\omega_{cs}}

\newcommand{\kpar}{k_\parallel}
\newcommand{\kperp}{k_\perp}
\newcommand{\kpars}{k_{\parallel,s}}

\newcommand{\pphi}{p_\phi}

\newcommand{\vinj}{\vb/\va}

\newcommand{\linj}{\lambda_0}
\newcommand{\dl}{\Delta\lambda}

\newcommand{\vpar}{v_\parallel}
\newcommand{\vpres}{v_{\parallel,\text{res}}}
\newcommand{\vpreslres}{v_{\parallel,res,\lres}}
\newcommand{\vdrift}{v_{\text{Dr}}}
\newcommand{\vperp}{v_\perp}
\newcommand{\vperpz}{v_{\perp,0}}
\newcommand{\lres}{\ell}

\newcommand{\J}[2]{\mathscr{J}_{#1}^{#2}}
\newcommand{\Jlm}{\J{\lres}{m}}
\newcommand{\Jlg}{\J{\lres}{G}}
\newcommand{\Jlc}{\J{\lres}{C}}

\newcommand{\Bres}{\eta}
\newcommand{\bres}{\Bres}

\newcommand{\Jl}{J_\lres}
\newcommand{\Jlprime}{\deriv{\Jl}{\flr}}

\newcommand{\W}{\mathcal{E}}

\newcommand{\Pb}{P_b}

\newcommand{\vB}{\vec{B}}

\newcommand{\vk}{\vec{k}}

\newcommand{\xinj}{x_0}
\newcommand{\dx}{\Delta x}

\newcommand{\fb}{f_0}

\newcommand{\omegacires}{\avg{\bar{\omega}_{ci}}}

\newcommand{\kratraw}{\kpar/\kperp}
\newcommand{\krat}{\abs{\kratraw}}
\newcommand{\rhob}{\rho_{\perp b}}

\newcommand{\zp}{\zeta}

\newcommand{\eqlab}[1]{\quad\text{#1}}
\newcommand{\CAElab}{\eqlab{CAE}}
\newcommand{\GAElab}{\eqlab{GAE}}

\newcommand{\ftail}{f_\text{tail}}

\newcommand{\Ksym}{K}
\newcommand{\Kij}{\Ksym_{ij}}
\newcommand{\Kxx}{\Ksym_{11}}
\newcommand{\Kxy}{\Ksym_{12}}
\newcommand{\Kyy}{\Ksym_{22}}

\newcommand{\flr}{\xi}
\newcommand{\bvar}{A^{-1}}

\newcommand{\avg}[1]{\left\langle #1 \right\rangle}
\renewcommand{\vec}[1]{\bm{#1}}

\newcommand{\abs}[1]{\left|#1\right|}

\newcommand{\defined}{\equiv}
\newcommand{\like}{\sim}

\newcommand{\ord}[1]{\mathcal{O}\left(#1\right)}
\newcommand{\plusord}[1]{\, + \, \ord{#1}}

\newcommand{\approptoinn}[2]{\mathrel{\vcenter{
  \offinterlineskip\halign{\hfil$##$\cr
    #1\propto\cr\noalign{\kern2pt}#1\sim\cr\noalign{\kern-2pt}}}}}
\newcommand{\appropto}{\mathpalette\approptoinn\relax}

\newcommand{\tto}{\text{ to }}

\newcommand\numberthis{\addtocounter{equation}{1}\tag{\theequation}}

\newcommand{\pderiv}[2]{\frac{\partial #1}{\partial #2}}

\newcommand{\deriv}[2]{\frac{d #1}{d #2}}

\newcommand{\figref}[1]{Fig.\xspace\ref{#1}}
\renewcommand{\eqref}[1]{Eq.\xspace\ref{#1}}
\newcommand{\secref}[1]{Sec.\xspace\ref{#1}}
\newcommand{\citeref}[1]{Ref.\xspace\onlinecite{#1}}
\newcommand{\appref}[1]{Appendix\xspace\ref{#1}}
\newcommand{\tabref}[1]{Table\xspace\ref{#1}}

\newcommand{\code}[1]{\texttt{#1}\xspace}
\newcommand{\HYM}{\code{HYM}}
\newcommand{\TRANSP}{\code{TRANSP}}
\newcommand{\NUBEAM}{\code{NUBEAM}}

\newcommand{\myname}{J.B. Lestz} 
\newcommand{\Elena}{E.V. Belova} 
\newcommand{\Nikolai}{N.N. Gorelenkov} 
\newcommand{\Neal}{N.A. Crocker} 
\newcommand{\Shawn}{S.X. Tang}

\newcommand{\PPPL}{Princeton Plasma Physics Lab, Princeton, NJ 08543, USA}
\newcommand{\Princeton}{Department of Astrophysical Sciences, Princeton University, Princeton, NJ 08543, USA}
\newcommand{\UCLA}{Department of Physics and Astronomy, University of California, Los Angeles, CA 90095, USA}

\newcommand{\Podesta}{Podest{\`a}}
\newcommand{\Fulop}{F{\"u}l{\"o}p}

\definecolor{darkgreen}{rgb}{0,0.5,0}

%% file: analytics_cyclotron_s1.0.bbl
\begin{thebibliography}{68}%
\makeatletter
\providecommand \@ifxundefined [1]{%
 \@ifx{#1\undefined}
}%
\providecommand \@ifnum [1]{%
 \ifnum #1\expandafter \@firstoftwo
 \else \expandafter \@secondoftwo
 \fi
}%
\providecommand \@ifx [1]{%
 \ifx #1\expandafter \@firstoftwo
 \else \expandafter \@secondoftwo
 \fi
}%
\providecommand \natexlab [1]{#1}%
\providecommand \enquote  [1]{``#1''}%
\providecommand \bibnamefont  [1]{#1}%
\providecommand \bibfnamefont [1]{#1}%
\providecommand \citenamefont [1]{#1}%
\providecommand \href@noop [0]{\@secondoftwo}%
\providecommand \href [0]{\begingroup \@sanitize@url \@href}%
\providecommand \@href[1]{\@@startlink{#1}\@@href}%
\providecommand \@@href[1]{\endgroup#1\@@endlink}%
\providecommand \@sanitize@url [0]{\catcode `\\12\catcode `\$12\catcode
  `\&12\catcode `\#12\catcode `\^12\catcode `\_12\catcode `\%12\relax}%
\providecommand \@@startlink[1]{}%
\providecommand \@@endlink[0]{}%
\providecommand \url  [0]{\begingroup\@sanitize@url \@url }%
\providecommand \@url [1]{\endgroup\@href {#1}{\urlprefix }}%
\providecommand \urlprefix  [0]{URL }%
\providecommand \Eprint [0]{\href }%
\providecommand \doibase [0]{http://dx.doi.org/}%
\providecommand \selectlanguage [0]{\@gobble}%
\providecommand \bibinfo  [0]{\@secondoftwo}%
\providecommand \bibfield  [0]{\@secondoftwo}%
\providecommand \translation [1]{[#1]}%
\providecommand \BibitemOpen [0]{}%
\providecommand \bibitemStop [0]{}%
\providecommand \bibitemNoStop [0]{.\EOS\space}%
\providecommand \EOS [0]{\spacefactor3000\relax}%
\providecommand \BibitemShut  [1]{\csname bibitem#1\endcsname}%
\let\auto@bib@innerbib\@empty
\bibitem [{\citenamefont {Fredrickson}\ \emph {et~al.}(2001)\citenamefont
  {Fredrickson}, \citenamefont {Gorelenkov}, \citenamefont {Cheng},
  \citenamefont {Bell}, \citenamefont {Darrow}, \citenamefont {Johnson},
  \citenamefont {Kaye}, \citenamefont {LeBlanc}, \citenamefont {Menard},
  \citenamefont {Kubota},\ and\ \citenamefont {Peebles}}]{Fredrickson2001PRL}%
  \BibitemOpen
  \bibfield  {author} {\bibinfo {author} {\bibfnamefont {E.~D.}\ \bibnamefont
  {Fredrickson}}, \bibinfo {author} {\bibfnamefont {N.}~\bibnamefont
  {Gorelenkov}}, \bibinfo {author} {\bibfnamefont {C.~Z.}\ \bibnamefont
  {Cheng}}, \bibinfo {author} {\bibfnamefont {R.}~\bibnamefont {Bell}},
  \bibinfo {author} {\bibfnamefont {D.}~\bibnamefont {Darrow}}, \bibinfo
  {author} {\bibfnamefont {D.}~\bibnamefont {Johnson}}, \bibinfo {author}
  {\bibfnamefont {S.}~\bibnamefont {Kaye}}, \bibinfo {author} {\bibfnamefont
  {B.}~\bibnamefont {LeBlanc}}, \bibinfo {author} {\bibfnamefont
  {J.}~\bibnamefont {Menard}}, \bibinfo {author} {\bibfnamefont
  {S.}~\bibnamefont {Kubota}}, \ and\ \bibinfo {author} {\bibfnamefont
  {W.}~\bibnamefont {Peebles}},\ }\href {\doibase
  10.1103/PhysRevLett.87.145001} {\bibfield  {journal} {\bibinfo  {journal}
  {Phys. Rev. Lett.}\ }\textbf {\bibinfo {volume} {87}},\ \bibinfo {pages}
  {145001} (\bibinfo {year} {2001})}\BibitemShut {NoStop}%
\bibitem [{\citenamefont {Fredrickson}\ \emph {et~al.}(2002)\citenamefont
  {Fredrickson}, \citenamefont {Gorelenkov}, \citenamefont {Cheng},
  \citenamefont {Bell}, \citenamefont {Darrow}, \citenamefont {Gates},
  \citenamefont {Johnson}, \citenamefont {Kaye}, \citenamefont {LeBlanc},
  \citenamefont {McCune}, \citenamefont {Menard}, \citenamefont {Roquemore},\
  and\ \citenamefont {Kubota}}]{Fredrickson2002POP}%
  \BibitemOpen
  \bibfield  {author} {\bibinfo {author} {\bibfnamefont {E.~D.}\ \bibnamefont
  {Fredrickson}}, \bibinfo {author} {\bibfnamefont {N.}~\bibnamefont
  {Gorelenkov}}, \bibinfo {author} {\bibfnamefont {C.~Z.}\ \bibnamefont
  {Cheng}}, \bibinfo {author} {\bibfnamefont {R.}~\bibnamefont {Bell}},
  \bibinfo {author} {\bibfnamefont {D.}~\bibnamefont {Darrow}}, \bibinfo
  {author} {\bibfnamefont {D.}~\bibnamefont {Gates}}, \bibinfo {author}
  {\bibfnamefont {D.}~\bibnamefont {Johnson}}, \bibinfo {author} {\bibfnamefont
  {S.}~\bibnamefont {Kaye}}, \bibinfo {author} {\bibfnamefont {B.}~\bibnamefont
  {LeBlanc}}, \bibinfo {author} {\bibfnamefont {D.}~\bibnamefont {McCune}},
  \bibinfo {author} {\bibfnamefont {J.}~\bibnamefont {Menard}}, \bibinfo
  {author} {\bibfnamefont {L.}~\bibnamefont {Roquemore}}, \ and\ \bibinfo
  {author} {\bibfnamefont {S.}~\bibnamefont {Kubota}},\ }\href {\doibase
  10.1063/1.1464542} {\bibfield  {journal} {\bibinfo  {journal} {Physics of
  Plasmas}\ }\textbf {\bibinfo {volume} {9}},\ \bibinfo {pages} {2069}
  (\bibinfo {year} {2002})}\BibitemShut {NoStop}%
\bibitem [{\citenamefont {Fredrickson}, \citenamefont {Gorelenkov},\ and\
  \citenamefont {Menard}(2004)}]{Fredrickson2004POP}%
  \BibitemOpen
  \bibfield  {author} {\bibinfo {author} {\bibfnamefont {E.~D.}\ \bibnamefont
  {Fredrickson}}, \bibinfo {author} {\bibfnamefont {N.~N.}\ \bibnamefont
  {Gorelenkov}}, \ and\ \bibinfo {author} {\bibfnamefont {J.}~\bibnamefont
  {Menard}},\ }\href {\doibase 10.1063/1.1760094} {\bibfield  {journal}
  {\bibinfo  {journal} {Physics of Plasmas}\ }\textbf {\bibinfo {volume}
  {11}},\ \bibinfo {pages} {3653} (\bibinfo {year} {2004})}\BibitemShut
  {NoStop}%
\bibitem [{\citenamefont {Fredrickson}\ \emph {et~al.}(2006)\citenamefont
  {Fredrickson}, \citenamefont {Bell}, \citenamefont {Darrow}, \citenamefont
  {Fu}, \citenamefont {Gorelenkov}, \citenamefont {LeBlanc}, \citenamefont
  {Medley}, \citenamefont {Menard}, \citenamefont {Park}, \citenamefont
  {Roquemore}, \citenamefont {Heidbrink}, \citenamefont {Sabbagh},
  \citenamefont {Stutman}, \citenamefont {Tritz}, \citenamefont {Crocker},
  \citenamefont {Kubota}, \citenamefont {Peebles}, \citenamefont {Lee},\ and\
  \citenamefont {Levinton}}]{Fredrickson2006POP}%
  \BibitemOpen
  \bibfield  {author} {\bibinfo {author} {\bibfnamefont {E.~D.}\ \bibnamefont
  {Fredrickson}}, \bibinfo {author} {\bibfnamefont {R.~E.}\ \bibnamefont
  {Bell}}, \bibinfo {author} {\bibfnamefont {D.~S.}\ \bibnamefont {Darrow}},
  \bibinfo {author} {\bibfnamefont {G.~Y.}\ \bibnamefont {Fu}}, \bibinfo
  {author} {\bibfnamefont {N.~N.}\ \bibnamefont {Gorelenkov}}, \bibinfo
  {author} {\bibfnamefont {B.~P.}\ \bibnamefont {LeBlanc}}, \bibinfo {author}
  {\bibfnamefont {S.~S.}\ \bibnamefont {Medley}}, \bibinfo {author}
  {\bibfnamefont {J.~E.}\ \bibnamefont {Menard}}, \bibinfo {author}
  {\bibfnamefont {H.}~\bibnamefont {Park}}, \bibinfo {author} {\bibfnamefont
  {A.~L.}\ \bibnamefont {Roquemore}}, \bibinfo {author} {\bibfnamefont {W.~W.}\
  \bibnamefont {Heidbrink}}, \bibinfo {author} {\bibfnamefont {S.~A.}\
  \bibnamefont {Sabbagh}}, \bibinfo {author} {\bibfnamefont {D.}~\bibnamefont
  {Stutman}}, \bibinfo {author} {\bibfnamefont {K.}~\bibnamefont {Tritz}},
  \bibinfo {author} {\bibfnamefont {N.~A.}\ \bibnamefont {Crocker}}, \bibinfo
  {author} {\bibfnamefont {S.}~\bibnamefont {Kubota}}, \bibinfo {author}
  {\bibfnamefont {W.}~\bibnamefont {Peebles}}, \bibinfo {author} {\bibfnamefont
  {K.~C.}\ \bibnamefont {Lee}}, \ and\ \bibinfo {author} {\bibfnamefont
  {F.~M.}\ \bibnamefont {Levinton}},\ }\href {\doibase 10.1063/1.2178788}
  {\bibfield  {journal} {\bibinfo  {journal} {Physics of Plasmas}\ }\textbf
  {\bibinfo {volume} {13}},\ \bibinfo {pages} {056109} (\bibinfo {year}
  {2006})}\BibitemShut {NoStop}%
\bibitem [{\citenamefont {Crocker}\ \emph {et~al.}(2011)\citenamefont
  {Crocker}, \citenamefont {Peebles}, \citenamefont {Kubota}, \citenamefont
  {Zhang}, \citenamefont {Bell}, \citenamefont {Fredrickson}, \citenamefont
  {Gorelenkov}, \citenamefont {LeBlanc}, \citenamefont {Menard}, \citenamefont
  {\Podesta}, \citenamefont {Sabbagh}, \citenamefont {Tritz},\ and\
  \citenamefont {Yuh}}]{Crocker2011PPCF}%
  \BibitemOpen
  \bibfield  {author} {\bibinfo {author} {\bibfnamefont {N.~A.}\ \bibnamefont
  {Crocker}}, \bibinfo {author} {\bibfnamefont {W.~A.}\ \bibnamefont
  {Peebles}}, \bibinfo {author} {\bibfnamefont {S.}~\bibnamefont {Kubota}},
  \bibinfo {author} {\bibfnamefont {J.}~\bibnamefont {Zhang}}, \bibinfo
  {author} {\bibfnamefont {R.~E.}\ \bibnamefont {Bell}}, \bibinfo {author}
  {\bibfnamefont {E.~D.}\ \bibnamefont {Fredrickson}}, \bibinfo {author}
  {\bibfnamefont {N.~N.}\ \bibnamefont {Gorelenkov}}, \bibinfo {author}
  {\bibfnamefont {B.~P.}\ \bibnamefont {LeBlanc}}, \bibinfo {author}
  {\bibfnamefont {J.~E.}\ \bibnamefont {Menard}}, \bibinfo {author}
  {\bibfnamefont {M.}~\bibnamefont {\Podesta}}, \bibinfo {author}
  {\bibfnamefont {S.~A.}\ \bibnamefont {Sabbagh}}, \bibinfo {author}
  {\bibfnamefont {K.}~\bibnamefont {Tritz}}, \ and\ \bibinfo {author}
  {\bibfnamefont {H.}~\bibnamefont {Yuh}},\ }\href
  {http://stacks.iop.org/0741-3335/53/i=10/a=105001} {\bibfield  {journal}
  {\bibinfo  {journal} {Plasma Physics and Controlled Fusion}\ }\textbf
  {\bibinfo {volume} {53}},\ \bibinfo {pages} {105001} (\bibinfo {year}
  {2011})}\BibitemShut {NoStop}%
\bibitem [{\citenamefont {Fredrickson}\ \emph {et~al.}(2012)\citenamefont
  {Fredrickson}, \citenamefont {Gorelenkov}, \citenamefont {Belova},
  \citenamefont {Crocker}, \citenamefont {Kubota}, \citenamefont {Kramer},
  \citenamefont {LeBlanc}, \citenamefont {Bell}, \citenamefont {\Podesta},
  \citenamefont {Yuh},\ and\ \citenamefont {Levinton}}]{Fredrickson2012NF}%
  \BibitemOpen
  \bibfield  {author} {\bibinfo {author} {\bibfnamefont {E.}~\bibnamefont
  {Fredrickson}}, \bibinfo {author} {\bibfnamefont {N.}~\bibnamefont
  {Gorelenkov}}, \bibinfo {author} {\bibfnamefont {E.}~\bibnamefont {Belova}},
  \bibinfo {author} {\bibfnamefont {N.}~\bibnamefont {Crocker}}, \bibinfo
  {author} {\bibfnamefont {S.}~\bibnamefont {Kubota}}, \bibinfo {author}
  {\bibfnamefont {G.}~\bibnamefont {Kramer}}, \bibinfo {author} {\bibfnamefont
  {B.}~\bibnamefont {LeBlanc}}, \bibinfo {author} {\bibfnamefont
  {R.}~\bibnamefont {Bell}}, \bibinfo {author} {\bibfnamefont {M.}~\bibnamefont
  {\Podesta}}, \bibinfo {author} {\bibfnamefont {H.}~\bibnamefont {Yuh}}, \
  and\ \bibinfo {author} {\bibfnamefont {F.}~\bibnamefont {Levinton}},\ }\href
  {http://stacks.iop.org/0029-5515/52/i=4/a=043001} {\bibfield  {journal}
  {\bibinfo  {journal} {Nuclear Fusion}\ }\textbf {\bibinfo {volume} {52}},\
  \bibinfo {pages} {043001} (\bibinfo {year} {2012})}\BibitemShut {NoStop}%
\bibitem [{\citenamefont {Fredrickson}\ \emph {et~al.}(2013)\citenamefont
  {Fredrickson}, \citenamefont {Gorelenkov}, \citenamefont {Podesta},
  \citenamefont {Bortolon}, \citenamefont {Crocker}, \citenamefont {Gerhardt},
  \citenamefont {Bell}, \citenamefont {Diallo}, \citenamefont {LeBlanc},
  \citenamefont {Levinton},\ and\ \citenamefont {Yuh}}]{Fredrickson2013POP}%
  \BibitemOpen
  \bibfield  {author} {\bibinfo {author} {\bibfnamefont {E.~D.}\ \bibnamefont
  {Fredrickson}}, \bibinfo {author} {\bibfnamefont {N.~N.}\ \bibnamefont
  {Gorelenkov}}, \bibinfo {author} {\bibfnamefont {M.}~\bibnamefont {Podesta}},
  \bibinfo {author} {\bibfnamefont {A.}~\bibnamefont {Bortolon}}, \bibinfo
  {author} {\bibfnamefont {N.~A.}\ \bibnamefont {Crocker}}, \bibinfo {author}
  {\bibfnamefont {S.~P.}\ \bibnamefont {Gerhardt}}, \bibinfo {author}
  {\bibfnamefont {R.~E.}\ \bibnamefont {Bell}}, \bibinfo {author}
  {\bibfnamefont {A.}~\bibnamefont {Diallo}}, \bibinfo {author} {\bibfnamefont
  {B.}~\bibnamefont {LeBlanc}}, \bibinfo {author} {\bibfnamefont {F.~M.}\
  \bibnamefont {Levinton}}, \ and\ \bibinfo {author} {\bibfnamefont
  {H.}~\bibnamefont {Yuh}},\ }\href {\doibase 10.1063/1.4801663} {\bibfield
  {journal} {\bibinfo  {journal} {Physics of Plasmas}\ }\textbf {\bibinfo
  {volume} {20}},\ \bibinfo {pages} {042112} (\bibinfo {year}
  {2013})}\BibitemShut {NoStop}%
\bibitem [{\citenamefont {Crocker}\ \emph {et~al.}(2013)\citenamefont
  {Crocker}, \citenamefont {Fredrickson}, \citenamefont {Gorelenkov},
  \citenamefont {Peebles}, \citenamefont {Kubota}, \citenamefont {Bell},
  \citenamefont {Diallo}, \citenamefont {LeBlanc}, \citenamefont {Menard},
  \citenamefont {\Podesta}, \citenamefont {Tritz},\ and\ \citenamefont
  {Yuh}}]{Crocker2013NF}%
  \BibitemOpen
  \bibfield  {author} {\bibinfo {author} {\bibfnamefont {N.}~\bibnamefont
  {Crocker}}, \bibinfo {author} {\bibfnamefont {E.}~\bibnamefont
  {Fredrickson}}, \bibinfo {author} {\bibfnamefont {N.}~\bibnamefont
  {Gorelenkov}}, \bibinfo {author} {\bibfnamefont {W.}~\bibnamefont {Peebles}},
  \bibinfo {author} {\bibfnamefont {S.}~\bibnamefont {Kubota}}, \bibinfo
  {author} {\bibfnamefont {R.}~\bibnamefont {Bell}}, \bibinfo {author}
  {\bibfnamefont {A.}~\bibnamefont {Diallo}}, \bibinfo {author} {\bibfnamefont
  {B.}~\bibnamefont {LeBlanc}}, \bibinfo {author} {\bibfnamefont
  {J.}~\bibnamefont {Menard}}, \bibinfo {author} {\bibfnamefont
  {M.}~\bibnamefont {\Podesta}}, \bibinfo {author} {\bibfnamefont
  {K.}~\bibnamefont {Tritz}}, \ and\ \bibinfo {author} {\bibfnamefont
  {H.}~\bibnamefont {Yuh}},\ }\href
  {http://stacks.iop.org/0029-5515/53/i=4/a=043017} {\bibfield  {journal}
  {\bibinfo  {journal} {Nuclear Fusion}\ }\textbf {\bibinfo {volume} {53}},\
  \bibinfo {pages} {043017} (\bibinfo {year} {2013})}\BibitemShut {NoStop}%
\bibitem [{\citenamefont {Crocker}\ \emph
  {et~al.}(2018{\natexlab{a}})\citenamefont {Crocker}, \citenamefont {Kubota},
  \citenamefont {Peebles}, \citenamefont {Rhodes}, \citenamefont {Fredrickson},
  \citenamefont {Belova}, \citenamefont {Diallo}, \citenamefont {LeBlanc},\
  and\ \citenamefont {Sabbagh}}]{Crocker2017NF}%
  \BibitemOpen
  \bibfield  {author} {\bibinfo {author} {\bibfnamefont {N.}~\bibnamefont
  {Crocker}}, \bibinfo {author} {\bibfnamefont {S.}~\bibnamefont {Kubota}},
  \bibinfo {author} {\bibfnamefont {W.}~\bibnamefont {Peebles}}, \bibinfo
  {author} {\bibfnamefont {T.}~\bibnamefont {Rhodes}}, \bibinfo {author}
  {\bibfnamefont {E.}~\bibnamefont {Fredrickson}}, \bibinfo {author}
  {\bibfnamefont {E.}~\bibnamefont {Belova}}, \bibinfo {author} {\bibfnamefont
  {A.}~\bibnamefont {Diallo}}, \bibinfo {author} {\bibfnamefont
  {B.}~\bibnamefont {LeBlanc}}, \ and\ \bibinfo {author} {\bibfnamefont
  {S.}~\bibnamefont {Sabbagh}},\ }\href
  {http://stacks.iop.org/0029-5515/58/i=1/a=016051} {\bibfield  {journal}
  {\bibinfo  {journal} {Nuclear Fusion}\ }\textbf {\bibinfo {volume} {58}},\
  \bibinfo {pages} {016051} (\bibinfo {year} {2018}{\natexlab{a}})}\BibitemShut
  {NoStop}%
\bibitem [{\citenamefont {Fredrickson}\ \emph {et~al.}(2018)\citenamefont
  {Fredrickson}, \citenamefont {Belova}, \citenamefont {Gorelenkov},
  \citenamefont {Podestà}, \citenamefont {Bell}, \citenamefont {Crocker},
  \citenamefont {Diallo}, \citenamefont {LeBlanc},\ and\ \citenamefont {the
  NSTX-U~team}}]{Fredrickson2018NF}%
  \BibitemOpen
  \bibfield  {author} {\bibinfo {author} {\bibfnamefont {E.}~\bibnamefont
  {Fredrickson}}, \bibinfo {author} {\bibfnamefont {E.}~\bibnamefont {Belova}},
  \bibinfo {author} {\bibfnamefont {N.}~\bibnamefont {Gorelenkov}}, \bibinfo
  {author} {\bibfnamefont {M.}~\bibnamefont {Podestà}}, \bibinfo {author}
  {\bibfnamefont {R.}~\bibnamefont {Bell}}, \bibinfo {author} {\bibfnamefont
  {N.}~\bibnamefont {Crocker}}, \bibinfo {author} {\bibfnamefont
  {A.}~\bibnamefont {Diallo}}, \bibinfo {author} {\bibfnamefont
  {B.}~\bibnamefont {LeBlanc}}, \ and\ \bibinfo {author} {\bibnamefont {the
  NSTX-U~team}},\ }\href {http://stacks.iop.org/0029-5515/58/i=8/a=082022}
  {\bibfield  {journal} {\bibinfo  {journal} {Nuclear Fusion}\ }\textbf
  {\bibinfo {volume} {58}},\ \bibinfo {pages} {082022} (\bibinfo {year}
  {2018})}\BibitemShut {NoStop}%
\bibitem [{\citenamefont {Fredrickson}\ \emph {et~al.}(2019)\citenamefont
  {Fredrickson}, \citenamefont {Gorelenkov}, \citenamefont {Bell},
  \citenamefont {Diallo}, \citenamefont {LeBlanc},\ and\ \citenamefont
  {Podestà}}]{Fredrickson2019POP}%
  \BibitemOpen
  \bibfield  {author} {\bibinfo {author} {\bibfnamefont {E.~D.}\ \bibnamefont
  {Fredrickson}}, \bibinfo {author} {\bibfnamefont {N.~N.}\ \bibnamefont
  {Gorelenkov}}, \bibinfo {author} {\bibfnamefont {R.~E.}\ \bibnamefont
  {Bell}}, \bibinfo {author} {\bibfnamefont {A.}~\bibnamefont {Diallo}},
  \bibinfo {author} {\bibfnamefont {B.~P.}\ \bibnamefont {LeBlanc}}, \ and\
  \bibinfo {author} {\bibfnamefont {M.}~\bibnamefont {Podestà}},\ }\href
  {\doibase 10.1063/1.5081047} {\bibfield  {journal} {\bibinfo  {journal}
  {Physics of Plasmas}\ }\textbf {\bibinfo {volume} {26}},\ \bibinfo {pages}
  {032111} (\bibinfo {year} {2019})}\BibitemShut {NoStop}%
\bibitem [{\citenamefont {Appel}\ \emph {et~al.}(2008)\citenamefont {Appel},
  \citenamefont {\Fulop}, \citenamefont {Hole}, \citenamefont {Smith},
  \citenamefont {Pinches}, \citenamefont {Vann},\ and\ \citenamefont {the
  MAST~team}}]{Appel2008PPCF}%
  \BibitemOpen
  \bibfield  {author} {\bibinfo {author} {\bibfnamefont {L.~C.}\ \bibnamefont
  {Appel}}, \bibinfo {author} {\bibfnamefont {T.}~\bibnamefont {\Fulop}},
  \bibinfo {author} {\bibfnamefont {M.~J.}\ \bibnamefont {Hole}}, \bibinfo
  {author} {\bibfnamefont {H.~M.}\ \bibnamefont {Smith}}, \bibinfo {author}
  {\bibfnamefont {S.~D.}\ \bibnamefont {Pinches}}, \bibinfo {author}
  {\bibfnamefont {R.~G.~L.}\ \bibnamefont {Vann}}, \ and\ \bibinfo {author}
  {\bibnamefont {the MAST~team}},\ }\href
  {http://stacks.iop.org/0741-3335/50/i=11/a=115011} {\bibfield  {journal}
  {\bibinfo  {journal} {Plasma Physics and Controlled Fusion}\ }\textbf
  {\bibinfo {volume} {50}},\ \bibinfo {pages} {115011} (\bibinfo {year}
  {2008})}\BibitemShut {NoStop}%
\bibitem [{\citenamefont {Sharapov}\ \emph {et~al.}(2014)\citenamefont
  {Sharapov}, \citenamefont {Lilley}, \citenamefont {Akers}, \citenamefont
  {Ayed}, \citenamefont {Cecconello}, \citenamefont {Cook}, \citenamefont
  {Cunningham},\ and\ \citenamefont {Verwichte}}]{Sharapov2014PP}%
  \BibitemOpen
  \bibfield  {author} {\bibinfo {author} {\bibfnamefont {S.~E.}\ \bibnamefont
  {Sharapov}}, \bibinfo {author} {\bibfnamefont {M.~K.}\ \bibnamefont
  {Lilley}}, \bibinfo {author} {\bibfnamefont {R.}~\bibnamefont {Akers}},
  \bibinfo {author} {\bibfnamefont {N.~B.}\ \bibnamefont {Ayed}}, \bibinfo
  {author} {\bibfnamefont {M.}~\bibnamefont {Cecconello}}, \bibinfo {author}
  {\bibfnamefont {J.~W.~S.}\ \bibnamefont {Cook}}, \bibinfo {author}
  {\bibfnamefont {G.}~\bibnamefont {Cunningham}}, \ and\ \bibinfo {author}
  {\bibfnamefont {E.}~\bibnamefont {Verwichte}},\ }\href {\doibase
  10.1063/1.4891322} {\bibfield  {journal} {\bibinfo  {journal} {Physics of
  Plasmas}\ }\textbf {\bibinfo {volume} {21}},\ \bibinfo {pages} {082501}
  (\bibinfo {year} {2014})}\BibitemShut {NoStop}%
\bibitem [{\citenamefont {McClements}\ and\ \citenamefont
  {Fredrickson}(2017)}]{McClements2017PPCF}%
  \BibitemOpen
  \bibfield  {author} {\bibinfo {author} {\bibfnamefont {K.~G.}\ \bibnamefont
  {McClements}}\ and\ \bibinfo {author} {\bibfnamefont {E.~D.}\ \bibnamefont
  {Fredrickson}},\ }\href {http://stacks.iop.org/0741-3335/59/i=5/a=053001}
  {\bibfield  {journal} {\bibinfo  {journal} {Plasma Physics and Controlled
  Fusion}\ }\textbf {\bibinfo {volume} {59}},\ \bibinfo {pages} {053001}
  (\bibinfo {year} {2017})}\BibitemShut {NoStop}%
\bibitem [{\citenamefont {Heidbrink}\ \emph {et~al.}(2006)\citenamefont
  {Heidbrink}, \citenamefont {Fredrickson}, \citenamefont {Gorelenkov},
  \citenamefont {Rhodes},\ and\ \citenamefont {Zeeland}}]{Heidbrink2006NF}%
  \BibitemOpen
  \bibfield  {author} {\bibinfo {author} {\bibfnamefont {W.}~\bibnamefont
  {Heidbrink}}, \bibinfo {author} {\bibfnamefont {E.}~\bibnamefont
  {Fredrickson}}, \bibinfo {author} {\bibfnamefont {N.}~\bibnamefont
  {Gorelenkov}}, \bibinfo {author} {\bibfnamefont {T.}~\bibnamefont {Rhodes}},
  \ and\ \bibinfo {author} {\bibfnamefont {M.~V.}\ \bibnamefont {Zeeland}},\
  }\href {http://stacks.iop.org/0029-5515/46/i=2/a=016} {\bibfield  {journal}
  {\bibinfo  {journal} {Nuclear Fusion}\ }\textbf {\bibinfo {volume} {46}},\
  \bibinfo {pages} {324} (\bibinfo {year} {2006})}\BibitemShut {NoStop}%
\bibitem [{\citenamefont {Tang}\ \emph {et~al.}(2018)\citenamefont {Tang},
  \citenamefont {Crocker}, \citenamefont {Carter}, \citenamefont {Thome},
  \citenamefont {Pinsker}, \citenamefont {Pace},\ and\ \citenamefont
  {Heidbrink}}]{Tang2018APS}%
  \BibitemOpen
  \bibfield  {author} {\bibinfo {author} {\bibfnamefont {S.}~\bibnamefont
  {Tang}}, \bibinfo {author} {\bibfnamefont {N.}~\bibnamefont {Crocker}},
  \bibinfo {author} {\bibfnamefont {T.}~\bibnamefont {Carter}}, \bibinfo
  {author} {\bibfnamefont {K.}~\bibnamefont {Thome}}, \bibinfo {author}
  {\bibfnamefont {R.}~\bibnamefont {Pinsker}}, \bibinfo {author} {\bibfnamefont
  {D.}~\bibnamefont {Pace}}, \ and\ \bibinfo {author} {\bibfnamefont
  {W.}~\bibnamefont {Heidbrink}},\ }in\ \href
  {http://meetings.aps.org/Meeting/DPP18/Session/NP11.111} {\emph {\bibinfo
  {booktitle} {$60^{th}$ APS DPP Meeting}}}\ (\bibinfo {address} {Portland,
  OR},\ \bibinfo {year} {2018})\BibitemShut {NoStop}%
\bibitem [{\citenamefont {Crocker}\ \emph
  {et~al.}(2018{\natexlab{b}})\citenamefont {Crocker}, \citenamefont {Barada},
  \citenamefont {Tang}, \citenamefont {Thome}, \citenamefont {Pace},
  \citenamefont {Pinsker}, \citenamefont {Heidbrink}, \citenamefont {Rhodes},\
  and\ \citenamefont {Haye}}]{Crocker2018APS}%
  \BibitemOpen
  \bibfield  {author} {\bibinfo {author} {\bibfnamefont {N.}~\bibnamefont
  {Crocker}}, \bibinfo {author} {\bibfnamefont {K.}~\bibnamefont {Barada}},
  \bibinfo {author} {\bibfnamefont {S.}~\bibnamefont {Tang}}, \bibinfo {author}
  {\bibfnamefont {K.}~\bibnamefont {Thome}}, \bibinfo {author} {\bibfnamefont
  {D.}~\bibnamefont {Pace}}, \bibinfo {author} {\bibfnamefont {R.}~\bibnamefont
  {Pinsker}}, \bibinfo {author} {\bibfnamefont {W.}~\bibnamefont {Heidbrink}},
  \bibinfo {author} {\bibfnamefont {T.}~\bibnamefont {Rhodes}}, \ and\ \bibinfo
  {author} {\bibfnamefont {R.~L.}\ \bibnamefont {Haye}},\ }in\ \href
  {http://meetings.aps.org/Meeting/DPP18/Session/NP11.110} {\emph {\bibinfo
  {booktitle} {$60^{th}$ APS DPP Meeting}}}\ (\bibinfo {address} {Portland,
  OR},\ \bibinfo {year} {2018})\BibitemShut {NoStop}%
\bibitem [{\citenamefont {Gorelenkova}\ and\ \citenamefont
  {Gorelenkov}(1998)}]{Gorelenkova1998POP}%
  \BibitemOpen
  \bibfield  {author} {\bibinfo {author} {\bibfnamefont {M.~V.}\ \bibnamefont
  {Gorelenkova}}\ and\ \bibinfo {author} {\bibfnamefont {N.~N.}\ \bibnamefont
  {Gorelenkov}},\ }\href {\doibase 10.1063/1.873133} {\bibfield  {journal}
  {\bibinfo  {journal} {Physics of Plasmas}\ }\textbf {\bibinfo {volume} {5}},\
  \bibinfo {pages} {4104} (\bibinfo {year} {1998})}\BibitemShut {NoStop}%
\bibitem [{\citenamefont {Kolesnichenko}\ \emph {et~al.}(1998)\citenamefont
  {Kolesnichenko}, \citenamefont {Fülöp}, \citenamefont {Lisak},\ and\
  \citenamefont {Anderson}}]{Kolesnichenko1998NF}%
  \BibitemOpen
  \bibfield  {author} {\bibinfo {author} {\bibfnamefont {Y.}~\bibnamefont
  {Kolesnichenko}}, \bibinfo {author} {\bibfnamefont {T.}~\bibnamefont
  {Fülöp}}, \bibinfo {author} {\bibfnamefont {M.}~\bibnamefont {Lisak}}, \
  and\ \bibinfo {author} {\bibfnamefont {D.}~\bibnamefont {Anderson}},\ }\href
  {http://stacks.iop.org/0029-5515/38/i=12/a=311} {\bibfield  {journal}
  {\bibinfo  {journal} {Nuclear Fusion}\ }\textbf {\bibinfo {volume} {38}},\
  \bibinfo {pages} {1871} (\bibinfo {year} {1998})}\BibitemShut {NoStop}%
\bibitem [{\citenamefont {Smith}\ \emph {et~al.}(2003)\citenamefont {Smith},
  \citenamefont {\Fulop}, \citenamefont {Lisak},\ and\ \citenamefont
  {Anderson}}]{Smith2003POP}%
  \BibitemOpen
  \bibfield  {author} {\bibinfo {author} {\bibfnamefont {H.}~\bibnamefont
  {Smith}}, \bibinfo {author} {\bibfnamefont {T.}~\bibnamefont {\Fulop}},
  \bibinfo {author} {\bibfnamefont {M.}~\bibnamefont {Lisak}}, \ and\ \bibinfo
  {author} {\bibfnamefont {D.}~\bibnamefont {Anderson}},\ }\href {\doibase
  10.1063/1.1566441} {\bibfield  {journal} {\bibinfo  {journal} {Physics of
  Plasmas}\ }\textbf {\bibinfo {volume} {10}},\ \bibinfo {pages} {1437}
  (\bibinfo {year} {2003})}\BibitemShut {NoStop}%
\bibitem [{\citenamefont {Smith}\ and\ \citenamefont
  {Verwichte}(2009)}]{Smith2009PPCF}%
  \BibitemOpen
  \bibfield  {author} {\bibinfo {author} {\bibfnamefont {H.~M.}\ \bibnamefont
  {Smith}}\ and\ \bibinfo {author} {\bibfnamefont {E.}~\bibnamefont
  {Verwichte}},\ }\href {http://stacks.iop.org/0741-3335/51/i=7/a=075001}
  {\bibfield  {journal} {\bibinfo  {journal} {Plasma Physics and Controlled
  Fusion}\ }\textbf {\bibinfo {volume} {51}},\ \bibinfo {pages} {075001}
  (\bibinfo {year} {2009})}\BibitemShut {NoStop}%
\bibitem [{\citenamefont {Smith}\ and\ \citenamefont
  {Fredrickson}(2017)}]{Smith2017PPCF}%
  \BibitemOpen
  \bibfield  {author} {\bibinfo {author} {\bibfnamefont {H.~M.}\ \bibnamefont
  {Smith}}\ and\ \bibinfo {author} {\bibfnamefont {E.~D.}\ \bibnamefont
  {Fredrickson}},\ }\href {http://stacks.iop.org/0741-3335/59/i=3/a=035007}
  {\bibfield  {journal} {\bibinfo  {journal} {Plasma Physics and Controlled
  Fusion}\ }\textbf {\bibinfo {volume} {59}},\ \bibinfo {pages} {035007}
  (\bibinfo {year} {2017})}\BibitemShut {NoStop}%
\bibitem [{\citenamefont {Mahajan}\ and\ \citenamefont
  {Ross}(1983)}]{Mahajan1983bPF}%
  \BibitemOpen
  \bibfield  {author} {\bibinfo {author} {\bibfnamefont {S.~M.}\ \bibnamefont
  {Mahajan}}\ and\ \bibinfo {author} {\bibfnamefont {D.~W.}\ \bibnamefont
  {Ross}},\ }\href {\doibase 10.1063/1.864445} {\bibfield  {journal} {\bibinfo
  {journal} {The Physics of Fluids}\ }\textbf {\bibinfo {volume} {26}},\
  \bibinfo {pages} {2561} (\bibinfo {year} {1983})}\BibitemShut {NoStop}%
\bibitem [{\citenamefont {Gorelenkov}, \citenamefont {Cheng},\ and\
  \citenamefont {Fredrickson}(2002)}]{Gorelenkov2002POP}%
  \BibitemOpen
  \bibfield  {author} {\bibinfo {author} {\bibfnamefont {N.~N.}\ \bibnamefont
  {Gorelenkov}}, \bibinfo {author} {\bibfnamefont {C.~Z.}\ \bibnamefont
  {Cheng}}, \ and\ \bibinfo {author} {\bibfnamefont {E.}~\bibnamefont
  {Fredrickson}},\ }\href {\doibase 10.1063/1.1492803} {\bibfield  {journal}
  {\bibinfo  {journal} {Physics of Plasmas}\ }\textbf {\bibinfo {volume} {9}},\
  \bibinfo {pages} {3483} (\bibinfo {year} {2002})}\BibitemShut {NoStop}%
\bibitem [{\citenamefont {Gorelenkov}\ \emph {et~al.}(2002)\citenamefont
  {Gorelenkov}, \citenamefont {Cheng}, \citenamefont {Fredrickson},
  \citenamefont {Belova}, \citenamefont {Gates}, \citenamefont {Kaye},
  \citenamefont {Kramer}, \citenamefont {Nazikian},\ and\ \citenamefont
  {White}}]{Gorelenkov2002NF}%
  \BibitemOpen
  \bibfield  {author} {\bibinfo {author} {\bibfnamefont {N.}~\bibnamefont
  {Gorelenkov}}, \bibinfo {author} {\bibfnamefont {C.}~\bibnamefont {Cheng}},
  \bibinfo {author} {\bibfnamefont {E.}~\bibnamefont {Fredrickson}}, \bibinfo
  {author} {\bibfnamefont {E.}~\bibnamefont {Belova}}, \bibinfo {author}
  {\bibfnamefont {D.}~\bibnamefont {Gates}}, \bibinfo {author} {\bibfnamefont
  {S.}~\bibnamefont {Kaye}}, \bibinfo {author} {\bibfnamefont {G.}~\bibnamefont
  {Kramer}}, \bibinfo {author} {\bibfnamefont {R.}~\bibnamefont {Nazikian}}, \
  and\ \bibinfo {author} {\bibfnamefont {R.}~\bibnamefont {White}},\ }\href
  {http://stacks.iop.org/0029-5515/42/i=8/a=306} {\bibfield  {journal}
  {\bibinfo  {journal} {Nuclear Fusion}\ }\textbf {\bibinfo {volume} {42}},\
  \bibinfo {pages} {977} (\bibinfo {year} {2002})}\BibitemShut {NoStop}%
\bibitem [{\citenamefont {Gorelenkov}\ \emph {et~al.}(2006)\citenamefont
  {Gorelenkov}, \citenamefont {Fredrickson}, \citenamefont {Heidbrink},
  \citenamefont {Crocker}, \citenamefont {Kubota},\ and\ \citenamefont
  {Peebles}}]{Gorelenkov2006NF}%
  \BibitemOpen
  \bibfield  {author} {\bibinfo {author} {\bibfnamefont {N.}~\bibnamefont
  {Gorelenkov}}, \bibinfo {author} {\bibfnamefont {E.}~\bibnamefont
  {Fredrickson}}, \bibinfo {author} {\bibfnamefont {W.}~\bibnamefont
  {Heidbrink}}, \bibinfo {author} {\bibfnamefont {N.}~\bibnamefont {Crocker}},
  \bibinfo {author} {\bibfnamefont {S.}~\bibnamefont {Kubota}}, \ and\ \bibinfo
  {author} {\bibfnamefont {W.}~\bibnamefont {Peebles}},\ }\href
  {http://stacks.iop.org/0029-5515/46/i=10/a=S10} {\bibfield  {journal}
  {\bibinfo  {journal} {Nuclear Fusion}\ }\textbf {\bibinfo {volume} {46}},\
  \bibinfo {pages} {S933} (\bibinfo {year} {2006})}\BibitemShut {NoStop}%
\bibitem [{\citenamefont {Heidbrink}(2008)}]{Heidbrink2008POP}%
  \BibitemOpen
  \bibfield  {author} {\bibinfo {author} {\bibfnamefont {W.~W.}\ \bibnamefont
  {Heidbrink}},\ }\href {\doibase 10.1063/1.2838239} {\bibfield  {journal}
  {\bibinfo  {journal} {Physics of Plasmas}\ }\textbf {\bibinfo {volume}
  {15}},\ \bibinfo {pages} {055501} (\bibinfo {year} {2008})}\BibitemShut
  {NoStop}%
\bibitem [{\citenamefont {Gorelenkov}, \citenamefont {Pinches},\ and\
  \citenamefont {Toi}(2014)}]{Gorelenkov2014NF}%
  \BibitemOpen
  \bibfield  {author} {\bibinfo {author} {\bibfnamefont {N.}~\bibnamefont
  {Gorelenkov}}, \bibinfo {author} {\bibfnamefont {S.}~\bibnamefont {Pinches}},
  \ and\ \bibinfo {author} {\bibfnamefont {K.}~\bibnamefont {Toi}},\ }\href
  {http://stacks.iop.org/0029-5515/54/i=12/a=125001} {\bibfield  {journal}
  {\bibinfo  {journal} {Nuclear Fusion}\ }\textbf {\bibinfo {volume} {54}},\
  \bibinfo {pages} {125001} (\bibinfo {year} {2014})}\BibitemShut {NoStop}%
\bibitem [{\citenamefont {Kolesnichenko}\ \emph {et~al.}(2007)\citenamefont
  {Kolesnichenko}, \citenamefont {Lutsenko}, \citenamefont {Weller},
  \citenamefont {Werner}, \citenamefont {Yakovenko}, \citenamefont {Geiger},\
  and\ \citenamefont {Fesenyuk}}]{Kolesnichenko2007POP}%
  \BibitemOpen
  \bibfield  {author} {\bibinfo {author} {\bibfnamefont {Y.~I.}\ \bibnamefont
  {Kolesnichenko}}, \bibinfo {author} {\bibfnamefont {V.~V.}\ \bibnamefont
  {Lutsenko}}, \bibinfo {author} {\bibfnamefont {A.}~\bibnamefont {Weller}},
  \bibinfo {author} {\bibfnamefont {A.}~\bibnamefont {Werner}}, \bibinfo
  {author} {\bibfnamefont {Y.~V.}\ \bibnamefont {Yakovenko}}, \bibinfo {author}
  {\bibfnamefont {J.}~\bibnamefont {Geiger}}, \ and\ \bibinfo {author}
  {\bibfnamefont {O.~P.}\ \bibnamefont {Fesenyuk}},\ }\href {\doibase
  10.1063/1.2789558} {\bibfield  {journal} {\bibinfo  {journal} {Physics of
  Plasmas}\ }\textbf {\bibinfo {volume} {14}},\ \bibinfo {pages} {102504}
  (\bibinfo {year} {2007})}\BibitemShut {NoStop}%
\bibitem [{\citenamefont {Appert}\ \emph {et~al.}(1982)\citenamefont {Appert},
  \citenamefont {Gruber}, \citenamefont {Troyuon},\ and\ \citenamefont
  {Vaclavik}}]{Appert1982PP}%
  \BibitemOpen
  \bibfield  {author} {\bibinfo {author} {\bibfnamefont {K.}~\bibnamefont
  {Appert}}, \bibinfo {author} {\bibfnamefont {R.}~\bibnamefont {Gruber}},
  \bibinfo {author} {\bibfnamefont {F.}~\bibnamefont {Troyuon}}, \ and\
  \bibinfo {author} {\bibfnamefont {J.}~\bibnamefont {Vaclavik}},\ }\href
  {http://stacks.iop.org/0032-1028/24/i=9/a=010} {\bibfield  {journal}
  {\bibinfo  {journal} {Plasma Physics}\ }\textbf {\bibinfo {volume} {24}},\
  \bibinfo {pages} {1147} (\bibinfo {year} {1982})}\BibitemShut {NoStop}%
\bibitem [{\citenamefont {Mahajan}, \citenamefont {Ross},\ and\ \citenamefont
  {Chen}(1983)}]{Mahajan1983PF}%
  \BibitemOpen
  \bibfield  {author} {\bibinfo {author} {\bibfnamefont {S.~M.}\ \bibnamefont
  {Mahajan}}, \bibinfo {author} {\bibfnamefont {D.~W.}\ \bibnamefont {Ross}}, \
  and\ \bibinfo {author} {\bibfnamefont {G.}~\bibnamefont {Chen}},\ }\href
  {\doibase 10.1063/1.864404} {\bibfield  {journal} {\bibinfo  {journal} {The
  Physics of Fluids}\ }\textbf {\bibinfo {volume} {26}},\ \bibinfo {pages}
  {2195} (\bibinfo {year} {1983})}\BibitemShut {NoStop}%
\bibitem [{\citenamefont {Mahajan}(1984)}]{Mahajan1984PF}%
  \BibitemOpen
  \bibfield  {author} {\bibinfo {author} {\bibfnamefont {S.~M.}\ \bibnamefont
  {Mahajan}},\ }\href {\doibase 10.1063/1.864878} {\bibfield  {journal}
  {\bibinfo  {journal} {The Physics of Fluids}\ }\textbf {\bibinfo {volume}
  {27}},\ \bibinfo {pages} {2238} (\bibinfo {year} {1984})}\BibitemShut
  {NoStop}%
\bibitem [{\citenamefont {Li}, \citenamefont {Mahajan},\ and\ \citenamefont
  {Ross}(1987)}]{Li1987PF}%
  \BibitemOpen
  \bibfield  {author} {\bibinfo {author} {\bibfnamefont {Y.~M.}\ \bibnamefont
  {Li}}, \bibinfo {author} {\bibfnamefont {S.~M.}\ \bibnamefont {Mahajan}}, \
  and\ \bibinfo {author} {\bibfnamefont {D.~W.}\ \bibnamefont {Ross}},\ }\href
  {\doibase 10.1063/1.866260} {\bibfield  {journal} {\bibinfo  {journal} {The
  Physics of Fluids}\ }\textbf {\bibinfo {volume} {30}},\ \bibinfo {pages}
  {1466} (\bibinfo {year} {1987})}\BibitemShut {NoStop}%
\bibitem [{\citenamefont {Fu}\ and\ \citenamefont {Dam}(1989)}]{Fu1989PF}%
  \BibitemOpen
  \bibfield  {author} {\bibinfo {author} {\bibfnamefont {G.~Y.}\ \bibnamefont
  {Fu}}\ and\ \bibinfo {author} {\bibfnamefont {J.~W.~V.}\ \bibnamefont
  {Dam}},\ }\href {\doibase 10.1063/1.859175} {\bibfield  {journal} {\bibinfo
  {journal} {Physics of Fluids B: Plasma Physics}\ }\textbf {\bibinfo {volume}
  {1}},\ \bibinfo {pages} {2404} (\bibinfo {year} {1989})}\BibitemShut
  {NoStop}%
\bibitem [{\citenamefont {Dam}, \citenamefont {Fu},\ and\ \citenamefont
  {Cheng}(1990)}]{VanDam1990FT}%
  \BibitemOpen
  \bibfield  {author} {\bibinfo {author} {\bibfnamefont {J.~V.}\ \bibnamefont
  {Dam}}, \bibinfo {author} {\bibfnamefont {G.}~\bibnamefont {Fu}}, \ and\
  \bibinfo {author} {\bibfnamefont {C.}~\bibnamefont {Cheng}},\ }\href
  {http://www.ans.org/pubs/journals/fst/a_29282} {\bibfield  {journal}
  {\bibinfo  {journal} {Fusion Technology}\ }\textbf {\bibinfo {volume} {18}},\
  \bibinfo {pages} {461} (\bibinfo {year} {1990})}\BibitemShut {NoStop}%
\bibitem [{\citenamefont {Lestz}\ \emph {et~al.}(rt 2)\citenamefont {Lestz},
  \citenamefont {Gorelenkov}, \citenamefont {Belova}, \citenamefont {Tang},\
  and\ \citenamefont {Crocker}}]{Lestz2019p2}%
  \BibitemOpen
  \bibfield  {author} {\bibinfo {author} {\bibfnamefont {J.}~\bibnamefont
  {Lestz}}, \bibinfo {author} {\bibfnamefont {N.}~\bibnamefont {Gorelenkov}},
  \bibinfo {author} {\bibfnamefont {E.}~\bibnamefont {Belova}}, \bibinfo
  {author} {\bibfnamefont {S.}~\bibnamefont {Tang}}, \ and\ \bibinfo {author}
  {\bibfnamefont {N.}~\bibnamefont {Crocker}},\ }\href@noop {} {\bibfield
  {journal} {\bibinfo  {journal} {Physics of Plasmas}\ } (\bibinfo {year}
  {2019, part 2})}\BibitemShut {NoStop}%
\bibitem [{\citenamefont {Stutman}\ \emph {et~al.}(2009)\citenamefont
  {Stutman}, \citenamefont {Delgado-Aparicio}, \citenamefont {Gorelenkov},
  \citenamefont {Finkenthal}, \citenamefont {Fredrickson}, \citenamefont
  {Kaye}, \citenamefont {Mazzucato},\ and\ \citenamefont
  {Tritz}}]{Stutman2009PRL}%
  \BibitemOpen
  \bibfield  {author} {\bibinfo {author} {\bibfnamefont {D.}~\bibnamefont
  {Stutman}}, \bibinfo {author} {\bibfnamefont {L.}~\bibnamefont
  {Delgado-Aparicio}}, \bibinfo {author} {\bibfnamefont {N.}~\bibnamefont
  {Gorelenkov}}, \bibinfo {author} {\bibfnamefont {M.}~\bibnamefont
  {Finkenthal}}, \bibinfo {author} {\bibfnamefont {E.}~\bibnamefont
  {Fredrickson}}, \bibinfo {author} {\bibfnamefont {S.}~\bibnamefont {Kaye}},
  \bibinfo {author} {\bibfnamefont {E.}~\bibnamefont {Mazzucato}}, \ and\
  \bibinfo {author} {\bibfnamefont {K.}~\bibnamefont {Tritz}},\ }\href
  {\doibase 10.1103/PhysRevLett.102.115002} {\bibfield  {journal} {\bibinfo
  {journal} {Phys. Rev. Lett.}\ }\textbf {\bibinfo {volume} {102}},\ \bibinfo
  {pages} {115002} (\bibinfo {year} {2009})}\BibitemShut {NoStop}%
\bibitem [{\citenamefont {Ren}\ \emph {et~al.}(2017)\citenamefont {Ren},
  \citenamefont {Belova}, \citenamefont {Gorelenkov}, \citenamefont
  {Guttenfelder}, \citenamefont {Kaye}, \citenamefont {Mazzucato},
  \citenamefont {Peterson}, \citenamefont {Smith}, \citenamefont {Stutman},
  \citenamefont {Tritz}, \citenamefont {Wang}, \citenamefont {Yuh},
  \citenamefont {Bell}, \citenamefont {Domier},\ and\ \citenamefont
  {LeBlanc}}]{Ren2017NF}%
  \BibitemOpen
  \bibfield  {author} {\bibinfo {author} {\bibfnamefont {Y.}~\bibnamefont
  {Ren}}, \bibinfo {author} {\bibfnamefont {E.}~\bibnamefont {Belova}},
  \bibinfo {author} {\bibfnamefont {N.}~\bibnamefont {Gorelenkov}}, \bibinfo
  {author} {\bibfnamefont {W.}~\bibnamefont {Guttenfelder}}, \bibinfo {author}
  {\bibfnamefont {S.}~\bibnamefont {Kaye}}, \bibinfo {author} {\bibfnamefont
  {E.}~\bibnamefont {Mazzucato}}, \bibinfo {author} {\bibfnamefont
  {J.}~\bibnamefont {Peterson}}, \bibinfo {author} {\bibfnamefont
  {D.}~\bibnamefont {Smith}}, \bibinfo {author} {\bibfnamefont
  {D.}~\bibnamefont {Stutman}}, \bibinfo {author} {\bibfnamefont
  {K.}~\bibnamefont {Tritz}}, \bibinfo {author} {\bibfnamefont
  {W.}~\bibnamefont {Wang}}, \bibinfo {author} {\bibfnamefont {H.}~\bibnamefont
  {Yuh}}, \bibinfo {author} {\bibfnamefont {R.}~\bibnamefont {Bell}}, \bibinfo
  {author} {\bibfnamefont {C.}~\bibnamefont {Domier}}, \ and\ \bibinfo {author}
  {\bibfnamefont {B.}~\bibnamefont {LeBlanc}},\ }\href
  {http://stacks.iop.org/0029-5515/57/i=7/a=072002} {\bibfield  {journal}
  {\bibinfo  {journal} {Nuclear Fusion}\ }\textbf {\bibinfo {volume} {57}},\
  \bibinfo {pages} {072002} (\bibinfo {year} {2017})}\BibitemShut {NoStop}%
\bibitem [{\citenamefont {Gorelenkov}\ \emph {et~al.}(2010)\citenamefont
  {Gorelenkov}, \citenamefont {Stutman}, \citenamefont {Tritz}, \citenamefont
  {Boozer}, \citenamefont {Delgado-Aparicio}, \citenamefont {Fredrickson},
  \citenamefont {Kaye},\ and\ \citenamefont {White}}]{Gorelenkov2010NF}%
  \BibitemOpen
  \bibfield  {author} {\bibinfo {author} {\bibfnamefont {N.}~\bibnamefont
  {Gorelenkov}}, \bibinfo {author} {\bibfnamefont {D.}~\bibnamefont {Stutman}},
  \bibinfo {author} {\bibfnamefont {K.}~\bibnamefont {Tritz}}, \bibinfo
  {author} {\bibfnamefont {A.}~\bibnamefont {Boozer}}, \bibinfo {author}
  {\bibfnamefont {L.}~\bibnamefont {Delgado-Aparicio}}, \bibinfo {author}
  {\bibfnamefont {E.}~\bibnamefont {Fredrickson}}, \bibinfo {author}
  {\bibfnamefont {S.}~\bibnamefont {Kaye}}, \ and\ \bibinfo {author}
  {\bibfnamefont {R.}~\bibnamefont {White}},\ }\href
  {http://stacks.iop.org/0029-5515/50/i=8/a=084012} {\bibfield  {journal}
  {\bibinfo  {journal} {Nuclear Fusion}\ }\textbf {\bibinfo {volume} {50}},\
  \bibinfo {pages} {084012} (\bibinfo {year} {2010})}\BibitemShut {NoStop}%
\bibitem [{\citenamefont {Kolesnichenko}, \citenamefont {Yakovenko},\ and\
  \citenamefont {Lutsenko}(2010)}]{Kolesnichenko2010PRL}%
  \BibitemOpen
  \bibfield  {author} {\bibinfo {author} {\bibfnamefont {Y.~I.}\ \bibnamefont
  {Kolesnichenko}}, \bibinfo {author} {\bibfnamefont {Y.~V.}\ \bibnamefont
  {Yakovenko}}, \ and\ \bibinfo {author} {\bibfnamefont {V.~V.}\ \bibnamefont
  {Lutsenko}},\ }\href {\doibase 10.1103/PhysRevLett.104.075001} {\bibfield
  {journal} {\bibinfo  {journal} {Phys. Rev. Lett.}\ }\textbf {\bibinfo
  {volume} {104}},\ \bibinfo {pages} {075001} (\bibinfo {year}
  {2010})}\BibitemShut {NoStop}%
\bibitem [{\citenamefont {Kolesnichenko}\ \emph {et~al.}(2010)\citenamefont
  {Kolesnichenko}, \citenamefont {Yakovenko}, \citenamefont {Lutsenko},
  \citenamefont {White},\ and\ \citenamefont {Weller}}]{Kolesnichenko2010NF}%
  \BibitemOpen
  \bibfield  {author} {\bibinfo {author} {\bibfnamefont {Y.}~\bibnamefont
  {Kolesnichenko}}, \bibinfo {author} {\bibfnamefont {Y.}~\bibnamefont
  {Yakovenko}}, \bibinfo {author} {\bibfnamefont {V.}~\bibnamefont {Lutsenko}},
  \bibinfo {author} {\bibfnamefont {R.}~\bibnamefont {White}}, \ and\ \bibinfo
  {author} {\bibfnamefont {A.}~\bibnamefont {Weller}},\ }\href
  {http://stacks.iop.org/0029-5515/50/i=8/a=084011} {\bibfield  {journal}
  {\bibinfo  {journal} {Nuclear Fusion}\ }\textbf {\bibinfo {volume} {50}},\
  \bibinfo {pages} {084011} (\bibinfo {year} {2010})}\BibitemShut {NoStop}%
\bibitem [{\citenamefont {Belova}\ \emph {et~al.}(2015)\citenamefont {Belova},
  \citenamefont {Gorelenkov}, \citenamefont {Fredrickson}, \citenamefont
  {Tritz},\ and\ \citenamefont {Crocker}}]{Belova2015PRL}%
  \BibitemOpen
  \bibfield  {author} {\bibinfo {author} {\bibfnamefont {E.~V.}\ \bibnamefont
  {Belova}}, \bibinfo {author} {\bibfnamefont {N.~N.}\ \bibnamefont
  {Gorelenkov}}, \bibinfo {author} {\bibfnamefont {E.~D.}\ \bibnamefont
  {Fredrickson}}, \bibinfo {author} {\bibfnamefont {K.}~\bibnamefont {Tritz}},
  \ and\ \bibinfo {author} {\bibfnamefont {N.~A.}\ \bibnamefont {Crocker}},\
  }\href {\doibase 10.1103/PhysRevLett.115.015001} {\bibfield  {journal}
  {\bibinfo  {journal} {Phys. Rev. Lett.}\ }\textbf {\bibinfo {volume} {115}},\
  \bibinfo {pages} {015001} (\bibinfo {year} {2015})}\BibitemShut {NoStop}%
\bibitem [{\citenamefont {Belova}\ \emph {et~al.}(2017)\citenamefont {Belova},
  \citenamefont {Gorelenkov}, \citenamefont {Crocker}, \citenamefont {Lestz},
  \citenamefont {Fredrickson}, \citenamefont {Tang},\ and\ \citenamefont
  {Tritz}}]{Belova2017POP}%
  \BibitemOpen
  \bibfield  {author} {\bibinfo {author} {\bibfnamefont {E.~V.}\ \bibnamefont
  {Belova}}, \bibinfo {author} {\bibfnamefont {N.~N.}\ \bibnamefont
  {Gorelenkov}}, \bibinfo {author} {\bibfnamefont {N.~A.}\ \bibnamefont
  {Crocker}}, \bibinfo {author} {\bibfnamefont {J.~B.}\ \bibnamefont {Lestz}},
  \bibinfo {author} {\bibfnamefont {E.~D.}\ \bibnamefont {Fredrickson}},
  \bibinfo {author} {\bibfnamefont {S.}~\bibnamefont {Tang}}, \ and\ \bibinfo
  {author} {\bibfnamefont {K.}~\bibnamefont {Tritz}},\ }\href {\doibase
  10.1063/1.4979278} {\bibfield  {journal} {\bibinfo  {journal} {Physics of
  Plasmas}\ }\textbf {\bibinfo {volume} {24}},\ \bibinfo {pages} {042505}
  (\bibinfo {year} {2017})}\BibitemShut {NoStop}%
\bibitem [{\citenamefont {Kolesnichenko}\ and\ \citenamefont
  {Tykhyy}(2018)}]{Kolesnichenko2018POP}%
  \BibitemOpen
  \bibfield  {author} {\bibinfo {author} {\bibfnamefont {Y.~I.}\ \bibnamefont
  {Kolesnichenko}}\ and\ \bibinfo {author} {\bibfnamefont {A.~V.}\ \bibnamefont
  {Tykhyy}},\ }\href {\doibase 10.1063/1.5048380} {\bibfield  {journal}
  {\bibinfo  {journal} {Physics of Plasmas}\ }\textbf {\bibinfo {volume}
  {25}},\ \bibinfo {pages} {102507} (\bibinfo {year} {2018})}\BibitemShut
  {NoStop}%
\bibitem [{\citenamefont {Kolesnichenko}, \citenamefont {Yakovenko},\ and\
  \citenamefont {Tyshchenko}(2018)}]{Kolesnichenko2018bPOP}%
  \BibitemOpen
  \bibfield  {author} {\bibinfo {author} {\bibfnamefont {Y.~I.}\ \bibnamefont
  {Kolesnichenko}}, \bibinfo {author} {\bibfnamefont {Y.~V.}\ \bibnamefont
  {Yakovenko}}, \ and\ \bibinfo {author} {\bibfnamefont {M.~H.}\ \bibnamefont
  {Tyshchenko}},\ }\href {\doibase 10.1063/1.5049543} {\bibfield  {journal}
  {\bibinfo  {journal} {Physics of Plasmas}\ }\textbf {\bibinfo {volume}
  {25}},\ \bibinfo {pages} {122508} (\bibinfo {year} {2018})}\BibitemShut
  {NoStop}%
\bibitem [{\citenamefont {Fredrickson}\ \emph {et~al.}(2017)\citenamefont
  {Fredrickson}, \citenamefont {Belova}, \citenamefont {Battaglia},
  \citenamefont {Bell}, \citenamefont {Crocker}, \citenamefont {Darrow},
  \citenamefont {Diallo}, \citenamefont {Gerhardt}, \citenamefont {Gorelenkov},
  \citenamefont {LeBlanc}, \citenamefont {\Podesta},\ and\ \citenamefont {the
  NSTX-U~team}}]{Fredrickson2017PRL}%
  \BibitemOpen
  \bibfield  {author} {\bibinfo {author} {\bibfnamefont {E.~D.}\ \bibnamefont
  {Fredrickson}}, \bibinfo {author} {\bibfnamefont {E.~V.}\ \bibnamefont
  {Belova}}, \bibinfo {author} {\bibfnamefont {D.~J.}\ \bibnamefont
  {Battaglia}}, \bibinfo {author} {\bibfnamefont {R.~E.}\ \bibnamefont {Bell}},
  \bibinfo {author} {\bibfnamefont {N.~A.}\ \bibnamefont {Crocker}}, \bibinfo
  {author} {\bibfnamefont {D.~S.}\ \bibnamefont {Darrow}}, \bibinfo {author}
  {\bibfnamefont {A.}~\bibnamefont {Diallo}}, \bibinfo {author} {\bibfnamefont
  {S.~P.}\ \bibnamefont {Gerhardt}}, \bibinfo {author} {\bibfnamefont {N.~N.}\
  \bibnamefont {Gorelenkov}}, \bibinfo {author} {\bibfnamefont {B.~P.}\
  \bibnamefont {LeBlanc}}, \bibinfo {author} {\bibfnamefont {M.}~\bibnamefont
  {\Podesta}}, \ and\ \bibinfo {author} {\bibnamefont {the NSTX-U~team}},\
  }\href {\doibase 10.1103/PhysRevLett.118.265001} {\bibfield  {journal}
  {\bibinfo  {journal} {Phys. Rev. Lett.}\ }\textbf {\bibinfo {volume} {118}},\
  \bibinfo {pages} {265001} (\bibinfo {year} {2017})}\BibitemShut {NoStop}%
\bibitem [{\citenamefont {Kaye}\ and\ \citenamefont {the
  NSTX-U~team}(pted)}]{Kaye2019NF}%
  \BibitemOpen
  \bibfield  {author} {\bibinfo {author} {\bibfnamefont {S.~M.}\ \bibnamefont
  {Kaye}}\ and\ \bibinfo {author} {\bibnamefont {the NSTX-U~team}},\ }\href
  {http://iopscience.iop.org/10.1088/1741-4326/ab023a} {\bibfield  {journal}
  {\bibinfo  {journal} {Nuclear Fusion}\ } (\bibinfo {year} {2019
  accepted})}\BibitemShut {NoStop}%
\bibitem [{\citenamefont {Belova}\ \emph {et~al.}(2019)\citenamefont {Belova},
  \citenamefont {Fredrickson}, \citenamefont {Lestz}, \citenamefont {Crocker},\
  and\ \citenamefont {the NSTX-U~team}}]{Belova2019POP}%
  \BibitemOpen
  \bibfield  {author} {\bibinfo {author} {\bibfnamefont {E.}~\bibnamefont
  {Belova}}, \bibinfo {author} {\bibfnamefont {E.}~\bibnamefont {Fredrickson}},
  \bibinfo {author} {\bibfnamefont {J.}~\bibnamefont {Lestz}}, \bibinfo
  {author} {\bibfnamefont {N.}~\bibnamefont {Crocker}}, \ and\ \bibinfo
  {author} {\bibnamefont {the NSTX-U~team}},\ }\href@noop {} {\bibfield
  {journal} {\bibinfo  {journal} {Physics of Plasmas}\ } (\bibinfo {year}
  {submitted 2019})}\BibitemShut {NoStop}%
\bibitem [{\citenamefont {Velikov}, \citenamefont {Kolesnichenko},\ and\
  \citenamefont {Oraevskii}(1968)}]{Belikov1968ZHETF}%
  \BibitemOpen
  \bibfield  {author} {\bibinfo {author} {\bibfnamefont {V.~S.}\ \bibnamefont
  {Velikov}}, \bibinfo {author} {\bibfnamefont {Y.~I.}\ \bibnamefont
  {Kolesnichenko}}, \ and\ \bibinfo {author} {\bibfnamefont {V.~N.}\
  \bibnamefont {Oraevskii}},\ }\href@noop {} {\bibfield  {journal} {\bibinfo
  {journal} {Zh. Eksp. Teor. Fiz.}\ }\textbf {\bibinfo {volume} {55}},\
  \bibinfo {pages} {2210} (\bibinfo {year} {1968})}\BibitemShut {NoStop}%
\bibitem [{\citenamefont {Velikov}, \citenamefont {Kolesnichenko},\ and\
  \citenamefont {Oraevskii}(1969)}]{Belikov1969JETP}%
  \BibitemOpen
  \bibfield  {author} {\bibinfo {author} {\bibfnamefont {V.~S.}\ \bibnamefont
  {Velikov}}, \bibinfo {author} {\bibfnamefont {Y.~I.}\ \bibnamefont
  {Kolesnichenko}}, \ and\ \bibinfo {author} {\bibfnamefont {V.~N.}\
  \bibnamefont {Oraevskii}},\ }\href
  {http://www.jetp.ac.ru/cgi-bin/e/index/e/28/6/p1172?a=list} {\bibfield
  {journal} {\bibinfo  {journal} {Journal of Experimental and Theoretical
  Physics}\ }\textbf {\bibinfo {volume} {28}},\ \bibinfo {pages} {1172}
  (\bibinfo {year} {1969})}\BibitemShut {NoStop}%
\bibitem [{\citenamefont {Timofeev}\ and\ \citenamefont
  {Pistunovich}(1970)}]{Timofeevv5}%
  \BibitemOpen
  \bibfield  {author} {\bibinfo {author} {\bibfnamefont {A.~V.}\ \bibnamefont
  {Timofeev}}\ and\ \bibinfo {author} {\bibfnamefont {V.}~\bibnamefont
  {Pistunovich}},\ }in\ \href@noop {} {\emph {\bibinfo {booktitle} {Reviews of
  Plasma Physics}}},\ Vol.~\bibinfo {volume} {5},\ \bibinfo {editor} {edited
  by\ \bibinfo {editor} {\bibfnamefont {M.~A.}\ \bibnamefont {Leontovich}}}\
  (\bibinfo  {publisher} {Consultants Bureau},\ \bibinfo {address} {New York},\
  \bibinfo {year} {1970})\ pp.\ \bibinfo {pages} {401 -- 445}\BibitemShut
  {NoStop}%
\bibitem [{\citenamefont {Mikhailovskii}(1975)}]{Mikhailovskiiv6}%
  \BibitemOpen
  \bibfield  {author} {\bibinfo {author} {\bibfnamefont {A.~B.}\ \bibnamefont
  {Mikhailovskii}},\ }in\ \href@noop {} {\emph {\bibinfo {booktitle} {Reviews
  of Plasma Physics}}},\ Vol.~\bibinfo {volume} {6},\ \bibinfo {editor} {edited
  by\ \bibinfo {editor} {\bibfnamefont {M.~A.}\ \bibnamefont {Leontovich}}}\
  (\bibinfo  {publisher} {Consultants Bureau},\ \bibinfo {address} {New York},\
  \bibinfo {year} {1975})\ pp.\ \bibinfo {pages} {77 -- 159}\BibitemShut
  {NoStop}%
\bibitem [{\citenamefont {Gorelenkov}\ and\ \citenamefont
  {Cheng}(1995{\natexlab{a}})}]{Gorelenkov1995POP}%
  \BibitemOpen
  \bibfield  {author} {\bibinfo {author} {\bibfnamefont {N.~N.}\ \bibnamefont
  {Gorelenkov}}\ and\ \bibinfo {author} {\bibfnamefont {C.~Z.}\ \bibnamefont
  {Cheng}},\ }\href {\doibase 10.1063/1.871281} {\bibfield  {journal} {\bibinfo
   {journal} {Physics of Plasmas}\ }\textbf {\bibinfo {volume} {2}},\ \bibinfo
  {pages} {1961} (\bibinfo {year} {1995}{\natexlab{a}})}\BibitemShut {NoStop}%
\bibitem [{\citenamefont {Gorelenkov}\ and\ \citenamefont
  {Cheng}(1995{\natexlab{b}})}]{Gorelenkov1995NF}%
  \BibitemOpen
  \bibfield  {author} {\bibinfo {author} {\bibfnamefont {N.}~\bibnamefont
  {Gorelenkov}}\ and\ \bibinfo {author} {\bibfnamefont {C.}~\bibnamefont
  {Cheng}},\ }\href {http://stacks.iop.org/0029-5515/35/i=12/a=I39} {\bibfield
  {journal} {\bibinfo  {journal} {Nuclear Fusion}\ }\textbf {\bibinfo {volume}
  {35}},\ \bibinfo {pages} {1743} (\bibinfo {year}
  {1995}{\natexlab{b}})}\BibitemShut {NoStop}%
\bibitem [{\citenamefont {Gorelenkov}\ and\ \citenamefont
  {Cheng}(2002)}]{Gorelenkov2002bNF}%
  \BibitemOpen
  \bibfield  {author} {\bibinfo {author} {\bibfnamefont {N.}~\bibnamefont
  {Gorelenkov}}\ and\ \bibinfo {author} {\bibfnamefont {C.}~\bibnamefont
  {Cheng}},\ }\href {http://stacks.iop.org/0029-5515/42/i=10/a=307} {\bibfield
  {journal} {\bibinfo  {journal} {Nuclear Fusion}\ }\textbf {\bibinfo {volume}
  {42}},\ \bibinfo {pages} {1216} (\bibinfo {year} {2002})}\BibitemShut
  {NoStop}%
\bibitem [{\citenamefont {Gorelenkov}\ \emph {et~al.}(2003)\citenamefont
  {Gorelenkov}, \citenamefont {Fredrickson}, \citenamefont {Belova},
  \citenamefont {Cheng}, \citenamefont {Gates}, \citenamefont {Kaye},\ and\
  \citenamefont {White}}]{Gorelenkov2003NF}%
  \BibitemOpen
  \bibfield  {author} {\bibinfo {author} {\bibfnamefont {N.}~\bibnamefont
  {Gorelenkov}}, \bibinfo {author} {\bibfnamefont {E.}~\bibnamefont
  {Fredrickson}}, \bibinfo {author} {\bibfnamefont {E.}~\bibnamefont {Belova}},
  \bibinfo {author} {\bibfnamefont {C.}~\bibnamefont {Cheng}}, \bibinfo
  {author} {\bibfnamefont {D.}~\bibnamefont {Gates}}, \bibinfo {author}
  {\bibfnamefont {S.}~\bibnamefont {Kaye}}, \ and\ \bibinfo {author}
  {\bibfnamefont {R.}~\bibnamefont {White}},\ }\href
  {http://stacks.iop.org/0029-5515/43/i=4/a=302} {\bibfield  {journal}
  {\bibinfo  {journal} {Nuclear Fusion}\ }\textbf {\bibinfo {volume} {43}},\
  \bibinfo {pages} {228} (\bibinfo {year} {2003})}\BibitemShut {NoStop}%
\bibitem [{\citenamefont {Kolesnichenko}, \citenamefont {White},\ and\
  \citenamefont {Yakovenko}(2006)}]{Kolesnichenko2006POP}%
  \BibitemOpen
  \bibfield  {author} {\bibinfo {author} {\bibfnamefont {Y.~I.}\ \bibnamefont
  {Kolesnichenko}}, \bibinfo {author} {\bibfnamefont {R.~B.}\ \bibnamefont
  {White}}, \ and\ \bibinfo {author} {\bibfnamefont {Y.~V.}\ \bibnamefont
  {Yakovenko}},\ }\href {\doibase 10.1063/1.2402129} {\bibfield  {journal}
  {\bibinfo  {journal} {Physics of Plasmas}\ }\textbf {\bibinfo {volume}
  {13}},\ \bibinfo {pages} {122503} (\bibinfo {year} {2006})}\BibitemShut
  {NoStop}%
\bibitem [{\citenamefont {Pankin}\ \emph {et~al.}(2004)\citenamefont {Pankin},
  \citenamefont {McCune}, \citenamefont {Andre}, \citenamefont {Bateman},\ and\
  \citenamefont {Kritz}}]{Pankin2004CPC}%
  \BibitemOpen
  \bibfield  {author} {\bibinfo {author} {\bibfnamefont {A.}~\bibnamefont
  {Pankin}}, \bibinfo {author} {\bibfnamefont {D.}~\bibnamefont {McCune}},
  \bibinfo {author} {\bibfnamefont {R.}~\bibnamefont {Andre}}, \bibinfo
  {author} {\bibfnamefont {G.}~\bibnamefont {Bateman}}, \ and\ \bibinfo
  {author} {\bibfnamefont {A.}~\bibnamefont {Kritz}},\ }\href {\doibase
  http://doi.org/10.1016/j.cpc.2003.11.002} {\bibfield  {journal} {\bibinfo
  {journal} {Computer Physics Communications}\ }\textbf {\bibinfo {volume}
  {159}},\ \bibinfo {pages} {157 } (\bibinfo {year} {2004})}\BibitemShut
  {NoStop}%
\bibitem [{\citenamefont {Goldston}\ \emph {et~al.}(1981)\citenamefont
  {Goldston}, \citenamefont {McCune}, \citenamefont {Towner}, \citenamefont
  {Davis}, \citenamefont {Hawryluk},\ and\ \citenamefont
  {Schmidt}}]{Goldston1982JCP}%
  \BibitemOpen
  \bibfield  {author} {\bibinfo {author} {\bibfnamefont {R.}~\bibnamefont
  {Goldston}}, \bibinfo {author} {\bibfnamefont {D.}~\bibnamefont {McCune}},
  \bibinfo {author} {\bibfnamefont {H.}~\bibnamefont {Towner}}, \bibinfo
  {author} {\bibfnamefont {S.}~\bibnamefont {Davis}}, \bibinfo {author}
  {\bibfnamefont {R.}~\bibnamefont {Hawryluk}}, \ and\ \bibinfo {author}
  {\bibfnamefont {G.}~\bibnamefont {Schmidt}},\ }\href {\doibase
  http://dx.doi.org/10.1016/0021-9991(81)90111-X} {\bibfield  {journal}
  {\bibinfo  {journal} {Journal of Computational Physics}\ }\textbf {\bibinfo
  {volume} {43}},\ \bibinfo {pages} {61 } (\bibinfo {year} {1981})}\BibitemShut
  {NoStop}%
\bibitem [{\citenamefont {Stix}(1975)}]{Stix1975NF}%
  \BibitemOpen
  \bibfield  {author} {\bibinfo {author} {\bibfnamefont {T.}~\bibnamefont
  {Stix}},\ }\href {http://stacks.iop.org/0029-5515/15/i=5/a=003} {\bibfield
  {journal} {\bibinfo  {journal} {Nuclear Fusion}\ }\textbf {\bibinfo {volume}
  {15}},\ \bibinfo {pages} {737} (\bibinfo {year} {1975})}\BibitemShut
  {NoStop}%
\bibitem [{\citenamefont {Lestz}, \citenamefont {Belova},\ and\ \citenamefont
  {Gorelenkov}(tion)}]{Lestz2019sim}%
  \BibitemOpen
  \bibfield  {author} {\bibinfo {author} {\bibfnamefont {J.}~\bibnamefont
  {Lestz}}, \bibinfo {author} {\bibfnamefont {E.}~\bibnamefont {Belova}}, \
  and\ \bibinfo {author} {\bibfnamefont {N.}~\bibnamefont {Gorelenkov}},\
  }\href@noop {} {\bibfield  {journal} {\bibinfo  {journal} {Physics of
  Plasmas}\ } (\bibinfo {year} {2019 in preparation})}\BibitemShut {NoStop}%
\bibitem [{\citenamefont {Belikov}, \citenamefont {Kolesnichenko},\ and\
  \citenamefont {White}(2003)}]{Belikov2003POP}%
  \BibitemOpen
  \bibfield  {author} {\bibinfo {author} {\bibfnamefont {V.~S.}\ \bibnamefont
  {Belikov}}, \bibinfo {author} {\bibfnamefont {Y.~I.}\ \bibnamefont
  {Kolesnichenko}}, \ and\ \bibinfo {author} {\bibfnamefont {R.~B.}\
  \bibnamefont {White}},\ }\href {\doibase http://dx.doi.org/10.1063/1.1625375}
  {\bibfield  {journal} {\bibinfo  {journal} {Physics of Plasmas}\ }\textbf
  {\bibinfo {volume} {10}},\ \bibinfo {pages} {4771} (\bibinfo {year}
  {2003})}\BibitemShut {NoStop}%
\bibitem [{\citenamefont {Belikov}, \citenamefont {Kolesnichenko},\ and\
  \citenamefont {White}(2004)}]{Belikov2004POP}%
  \BibitemOpen
  \bibfield  {author} {\bibinfo {author} {\bibfnamefont {V.~S.}\ \bibnamefont
  {Belikov}}, \bibinfo {author} {\bibfnamefont {Y.~I.}\ \bibnamefont
  {Kolesnichenko}}, \ and\ \bibinfo {author} {\bibfnamefont {R.~B.}\
  \bibnamefont {White}},\ }\href {\doibase http://dx.doi.org/10.1063/1.1809121}
  {\bibfield  {journal} {\bibinfo  {journal} {Physics of Plasmas}\ }\textbf
  {\bibinfo {volume} {11}},\ \bibinfo {pages} {5409} (\bibinfo {year}
  {2004})}\BibitemShut {NoStop}%
\bibitem [{\citenamefont {Belova}, \citenamefont {Gorelenkov},\ and\
  \citenamefont {Cheng}(2003)}]{Belova2003POP}%
  \BibitemOpen
  \bibfield  {author} {\bibinfo {author} {\bibfnamefont {E.~V.}\ \bibnamefont
  {Belova}}, \bibinfo {author} {\bibfnamefont {N.~N.}\ \bibnamefont
  {Gorelenkov}}, \ and\ \bibinfo {author} {\bibfnamefont {C.~Z.}\ \bibnamefont
  {Cheng}},\ }\href {\doibase http://dx.doi.org/10.1063/1.1592155} {\bibfield
  {journal} {\bibinfo  {journal} {Physics of Plasmas}\ }\textbf {\bibinfo
  {volume} {10}},\ \bibinfo {pages} {3240} (\bibinfo {year}
  {2003})}\BibitemShut {NoStop}%
\bibitem [{\citenamefont {Lestz}, \citenamefont {Belova},\ and\ \citenamefont
  {Gorelenkov}(2018{\natexlab{a}})}]{Lestz2018POP}%
  \BibitemOpen
  \bibfield  {author} {\bibinfo {author} {\bibfnamefont {J.~B.}\ \bibnamefont
  {Lestz}}, \bibinfo {author} {\bibfnamefont {E.~V.}\ \bibnamefont {Belova}}, \
  and\ \bibinfo {author} {\bibfnamefont {N.~N.}\ \bibnamefont {Gorelenkov}},\
  }\href {\doibase 10.1063/1.4998602} {\bibfield  {journal} {\bibinfo
  {journal} {Physics of Plasmas}\ }\textbf {\bibinfo {volume} {25}},\ \bibinfo
  {pages} {042508} (\bibinfo {year} {2018}{\natexlab{a}})}\BibitemShut
  {NoStop}%
\bibitem [{Ben(1978)}]{BenderOrszagStationaryPhase}%
  \BibitemOpen
  \enquote {\bibinfo {title} {Advanced mathematical methods for scientists and
  engineers: Asymptotic methods and perturbation theory},}\ \ (\bibinfo
  {publisher} {McGraw-Hill},\ \bibinfo {year} {1978})\ Chap.\ \bibinfo
  {chapter} {6.5 Method of Stationary Phase}\BibitemShut {NoStop}%
\bibitem [{\citenamefont {Lestz}, \citenamefont {Belova},\ and\ \citenamefont
  {Gorelenkov}(2018{\natexlab{b}})}]{Lestz2018APS}%
  \BibitemOpen
  \bibfield  {author} {\bibinfo {author} {\bibfnamefont {J.}~\bibnamefont
  {Lestz}}, \bibinfo {author} {\bibfnamefont {E.}~\bibnamefont {Belova}}, \
  and\ \bibinfo {author} {\bibfnamefont {N.}~\bibnamefont {Gorelenkov}},\ }in\
  \href {http://meetings.aps.org/Meeting/DPP18/Session/TP11.00096} {\emph
  {\bibinfo {booktitle} {$60^{th}$ APS DPP Meeting}}}\ (\bibinfo {address}
  {Portland, OR},\ \bibinfo {year} {2018})\BibitemShut {NoStop}%
\bibitem [{\citenamefont {Tang}\ \emph {et~al.}(2017)\citenamefont {Tang},
  \citenamefont {Crocker}, \citenamefont {Carter}, \citenamefont {Fredrickson},
  \citenamefont {Gorelenkov},\ and\ \citenamefont
  {Guttenfelder}}]{Tang2017TTF}%
  \BibitemOpen
  \bibfield  {author} {\bibinfo {author} {\bibfnamefont {S.}~\bibnamefont
  {Tang}}, \bibinfo {author} {\bibfnamefont {N.}~\bibnamefont {Crocker}},
  \bibinfo {author} {\bibfnamefont {T.}~\bibnamefont {Carter}}, \bibinfo
  {author} {\bibfnamefont {E.}~\bibnamefont {Fredrickson}}, \bibinfo {author}
  {\bibfnamefont {N.}~\bibnamefont {Gorelenkov}}, \ and\ \bibinfo {author}
  {\bibfnamefont {W.}~\bibnamefont {Guttenfelder}},\ }in\ \href@noop {} {\emph
  {\bibinfo {booktitle} {2017 US/EU Transport Task Force}}}\ (\bibinfo
  {address} {Williamsburg, VA},\ \bibinfo {year} {2017})\BibitemShut {NoStop}%
\end{thebibliography}%
